\documentclass[12pt]{article}
\usepackage{amsmath,amsthm,latexsym,amssymb,amsfonts,epsfig}

\makeatletter
\@addtoreset{equation}{section}
\makeatother

\def\be{\begin{equation}}
\def\ee{\end{equation}}
\def\ba{\begin{eqnarray}}
\def\ea{\end{eqnarray}}
\newcommand{\lp}{\ell_p}

\newcommand{\sst}{\scriptscriptstyle}

\newcommand{\op}[1]{\widehat{#1}}
\newcommand{\wt}[1]{\widetilde{#1}}
\newcommand{\lb}{\left}
\newcommand{\rb}{\right}

\newcommand{\vt}{\tilde{v}}

\newcommand{\mn}{m_{\sst 0}}
\newcommand{\mnt}{\wt{m}_{\sst 0}}
\newcommand{\nn}{n_{\sst 0}}
\newcommand{\nnt}{\wt{n}_{\sst 0}}
\newcommand{\elln}{\ell_{\sst 0}}

\newcommand{\Jn}{J_{\sst 0}}
\newcommand{\Jnt}{\wt{J}_{\sst 0}}
\newcommand{\Kn}{K_{\sst 0}}
\newcommand{\Knt}{\wt{K}_{\sst 0}}
\newcommand{\In}{I_{\sst 0}}
\newcommand{\Int}{\wt{I}_{\sst 0}}
\newcommand{\sn}{\sigma_{\sst 0}}
\newcommand{\snt}{\wt{\sigma}_{\sst 0}}

\newcommand{\n}[4]{n_{#1 #2 #3 #4}}
\newcommand{\hol}[4]{\op{h}_{#1 #2 #3 #4}}
\newcommand{\holin}[4]{\op{h}^{-1}_{#1 #2 #3 #4}}

\newcommand{\holloop}[5]{\op{h}_{\alpha_{#1 #2 #3 #4 #5}}}

\newcommand{\Psigen}{\Psi^{t}_{\{g,J,\sigma,j,L\}}}

\newcommand{\Psigeng}{\Psi^{t}_{\gamma,m}}

\newcommand{\hgen}{\hat{h}_{J \sigma j v}}

\newcommand{\pmm}{(p)^-}

\newcommand{\ngen}{n_{J \sigma j \tilde{v}}}

\newcommand{\qmm}{(q^{-1})^-}

\newcommand{\phigen}{\varphi_{J \sigma j \tilde{v}}}

\newcommand{\Del}{\Delta(\In,\Int,\Jn,\Jnt,\sn,\snt,\mn,\mnt,v,J,\sigma,j,\vt)}

\newcommand{\Ygen}{\op{X}_{J\sigma j v}}
\newcommand{\Y}[4]{\op{X}_{#1 #2 #3 #4}}

\def\Nl{{\mathchoice
{\setbox0=\hbox{$\displaystyle\rm N$}\hbox{\hbox to0pt
{\kern0.4\wd0\vrule height0.9\ht0\hss}\box0}}
{\setbox0=\hbox{$\textstyle\rm N$}\hbox{\hbox to0pt
{\kern0.4\wd0\vrule height0.9\ht0\hss}\box0}}
{\setbox0=\hbox{$\scriptstyle\rm N$}\hbox{\hbox to0pt
{\kern0.4\wd0\vrule height0.9\ht0\hss}\box0}}
{\setbox0=\hbox{$\scriptscriptstyle\rm N$}\hbox{\hbox to0pt
{\kern0.4\wd0\vrule height0.9\ht0\hss}\box0}}}}
\def\Zl{{\mathchoice
{\setbox0=\hbox{$\displaystyle\rm Z$}\hbox{\hbox to0pt
{\kern0.4\wd0\vrule height0.9\ht0\hss}\box0}}
{\setbox0=\hbox{$\textstyle\rm Z$}\hbox{\hbox to0pt
{\kern0.4\wd0\vrule height0.9\ht0\hss}\box0}}
{\setbox0=\hbox{$\scriptstyle\rm Z$}\hbox{\hbox to0pt
{\kern0.4\wd0\vrule height0.9\ht0\hss}\box0}}
{\setbox0=\hbox{$\scriptscriptstyle\rm Z$}\hbox{\hbox to0pt
{\kern0.4\wd0\vrule height0.9\ht0\hss}\box0}}}}
\def\Ql{{\mathchoice
{\setbox0=\hbox{$\displaystyle\rm Q$}\hbox{\hbox to0pt
{\kern0.4\wd0\vrule height0.9\ht0\hss}\box0}}
{\setbox0=\hbox{$\textstyle\rm Q$}\hbox{\hbox to0pt
{\kern0.4\wd0\vrule height0.9\ht0\hss}\box0}}
{\setbox0=\hbox{$\scriptstyle\rm Q$}\hbox{\hbox to0pt
{\kern0.4\wd0\vrule height0.9\ht0\hss}\box0}}
{\setbox0=\hbox{$\scriptscriptstyle\rm Q$}\hbox{\hbox to0pt
{\kern0.4\wd0\vrule height0.9\ht0\hss}\box0}}}}
\def\Rl{{\mathchoice
{\setbox0=\hbox{$\displaystyle\rm R$}\hbox{\hbox to0pt
{\kern0.4\wd0\vrule height0.9\ht0\hss}\box0}}
{\setbox0=\hbox{$\textstyle\rm R$}\hbox{\hbox to0pt
{\kern0.4\wd0\vrule height0.9\ht0\hss}\box0}}
{\setbox0=\hbox{$\scriptstyle\rm R$}\hbox{\hbox to0pt
{\kern0.4\wd0\vrule height0.9\ht0\hss}\box0}}
{\setbox0=\hbox{$\scriptscriptstyle\rm R$}\hbox{\hbox to0pt
{\kern0.4\wd0\vrule height0.9\ht0\hss}\box0}}}}
\def\Cl{{\mathchoice
{\setbox0=\hbox{$\displaystyle\rm C$}\hbox{\hbox to0pt
{\kern0.4\wd0\vrule height0.9\ht0\hss}\box0}}
{\setbox0=\hbox{$\textstyle\rm C$}\hbox{\hbox to0pt
{\kern0.4\wd0\vrule height0.9\ht0\hss}\box0}}
{\setbox0=\hbox{$\scriptstyle\rm C$}\hbox{\hbox to0pt
{\kern0.4\wd0\vrule height0.9\ht0\hss}\box0}}
{\setbox0=\hbox{$\scriptscriptstyle\rm C$}\hbox{\hbox to0pt
{\kern0.4\wd0\vrule height0.9\ht0\hss}\box0}}}}
\def\Hl{{\mathchoice
{\setbox0=\hbox{$\displaystyle\rm H$}\hbox{\hbox to0pt
{\kern0.4\wd0\vrule height0.9\ht0\hss}\box0}}
{\setbox0=\hbox{$\textstyle\rm H$}\hbox{\hbox to0pt
{\kern0.4\wd0\vrule height0.9\ht0\hss}\box0}}
{\setbox0=\hbox{$\scriptstyle\rm H$}\hbox{\hbox to0pt
{\kern0.4\wd0\vrule height0.9\ht0\hss}\box0}}
{\setbox0=\hbox{$\scriptscriptstyle\rm H$}\hbox{\hbox to0pt
{\kern0.4\wd0\vrule height0.9\ht0\hss}\box0}}}}
\def\Ol{{\mathchoice
{\setbox0=\hbox{$\displaystyle\rm O$}\hbox{\hbox to0pt
{\kern0.4\wd0\vrule height0.9\ht0\hss}\box0}}
{\setbox0=\hbox{$\textstyle\rm O$}\hbox{\hbox to0pt
{\kern0.4\wd0\vrule height0.9\ht0\hss}\box0}}
{\setbox0=\hbox{$\scriptstyle\rm O$}\hbox{\hbox to0pt
{\kern0.4\wd0\vrule height0.9\ht0\hss}\box0}}
{\setbox0=\hbox{$\scriptscriptstyle\rm O$}\hbox{\hbox to0pt
{\kern0.4\wd0\vrule height0.9\ht0\hss}\box0}}}}
\DeclareMathOperator{\MC}{\boldsymbol{\mathsf{M}}}
\DeclareMathOperator{\MCO}{\boldsymbol{\widehat{\mathsf{M}}}}

\DeclareMathOperator{\sgn}{sgn}

\title{{\sf Algebraic Quantum Gravity (AQG)\\ I. Conceptual Setup}}
\author{\sf 
K. Giesel\thanks{{\sf gieskri@aei.mpg.de, kgiesel@perimeterinstitute.ca}} ~~ and ~ 
T. 
Thiemann\thanks{{\sf 
thiemann@aei.mpg.de,tthiemann@perimeterinstitute.ca}}\\
\\
{\sf MPI f. Gravitationsphysik, Albert-Einstein-Institut,} \\
           {\sf Am M\"uhlenberg 1, 14476 Potsdam, Germany}\\
\\
{\sf and}\\
\\
{\sf Perimeter Institute for Theoretical Physics,}\\ 
{\sf 31 Caroline Street N, Waterloo, ON N2L 2Y5, Canada}}
\date{{\small\sf Preprint AEI-2006-058}}

\begin{document}

\maketitle

\begin{abstract}
{\sf
We introduce a new top down approach to canonical quantum gravity,
called Algebraic Quantum Gravity (AQG):
The quantum kinematics of AQG is determined by an abstract 
$*-$algebra generated by a countable set of elementary operators 
labelled by an algebraic graph. The quantum dynamics of AQG is governed 
by a single Master Constraint operator. 
While AQG is inspired by Loop Quantum Gravity
(LQG), it differs drastically from it because in AQG there is 
fundamentally no topology or differential structure. A natural Hilbert 
space representation  
acquires the structure of an infinite tensor product (ITP) whose separable 
strong equivalence class Hilbert subspaces (sectors) are left invariant 
by the quantum dynamics.

The missing information 
about the topology and differential structure of the spacetime manifold as 
well as about the background metric to be approximated
is supplied by coherent states. Given such data,
the corresponding coherent state defines a sector in the ITP
which can be identified with a usual QFT on the given manifold 
and background.
Thus, AQG contains QFT on all curved spacetimes at once, 
possibly has something to say about topology change and 
provides the contact with the familiar low energy physics. In 
particular, in two companion papers we 
develop semiclassical perturbation theory for AQG and LQG and thereby 
show that the theory admits a semiclassical 
limit whose infinitesimal gauge symmetry agrees with that of General 
Relativity.

In AQG everything is computable with sufficient precision and no UV 
divergences arise due 
to the background independence of the fundamental combinatorial 
structure. 
Hence, in contrast to lattice gauge theory on a background metric, 
no continuum limit has to be taken, there simply is no lattice regulator 
that must be sent to zero. 

}
\end{abstract}

\newpage

\section{Introduction}
\label{s1}

The present paper introduces a new conceptual framework for canonical 
quantum gravity resulting in a novel top to bottom approach. To justify 
it, a rather complex reasoning is required based on the current status 
of the quantum dynamics of Loop Quantum Gravity (LQG). 
Therefore we will devote quite some space in this introduction to 
make the motivations, concepts and techniques clear and in order to show 
how this theory differs from the more traditional framework.

\subsection{Anomalies and the Semiclassical Analysis of LQG}
\label{s1.1}
 
Loop Quantum Gravity (LQG) has advanced in recent years to one of the 
major candidates for a theory of quantum gravity. See \cite{books} for 
books and \cite{reviews} for recent reviews on the subject. The theory 
has a mathematically rigorous basis of the quantum kinematics 
\cite{AI,AL,ALMMT} and there is a mathematically well defined formulation 
of the quantum dynamics \cite{QSD}. However, one problem has remained  
unsettled so far within LQG: The demonstration that the theory has General 
Relativity as its semiclassical limit. Related to this, so far it was not 
revealed that the algebra of the quantum constraints, while free of 
anomalies, mimics the algebra of the classical constraints.

The reason for this so far elusive evidence has a complicated but clear
technical reason and in what follows we will try to explain it in some 
detail:\\ 
In the current setup, LQG is 
formulated in terms of gauge field variables, that is, non -- Abelean 
electric fluxes and magnetic holonomies, just like in lattice gauge 
theory. The corresponding surfaces and curves are embedded into a 
spatial manifold $\sigma$ of given topology. These define an abstract 
$^\ast-$Poisson algebra. Using the physically well motivated condition of 
spatial diffeomorphism invariance, one can show that there is only one 
unitary equivalence class of cyclic representations of this holonomy -- 
flux algebra \cite{LOST,Fleischhack}. Thus, the {\it kinematical} 
framework of LQG is rather tight and well under control.  

The unique (up to unitary equivalence) Hilbert space $\cal H$ can be 
realised as 
the closure of the finite linear combinations of cylindrical functions.
A cylindrical function is a complex valued, square integrable (with 
respect to a certain measure) function of holonomies 
along the edges of some finite graph and all finite graphs embedded 
into $\sigma$ are allowed. 
Thus, in contrast to lattice gauge theory, the lattice is not fixed, 
rather, all lattices (or graphs) are considered simultaneously which is 
why LQG is a continuum rather than a lattice theory. 

The problem with establishing the semiclassical limit of LQG has to do 
with the quantum {\it dynamics}:\\
There is a natural action of the spatial diffeomorphism 
group Diff$(\sigma)$ on this Hilbert space which simply consists in 
mapping graphs to their images under the given diffeomorphism. 
This action is a unitary representation of Diff$(\sigma)$ and therefore 
the spatial diffeomorphisms are represented without anomalies. 
However, the action is not weakly continuous. This means that the
infinitesimal generators of Diff$(\sigma)$, that is, the Lie algebra
diff$(\sigma)$, cannot be defined on $\cal H$. In contrast, 
the infinite number of Hamiltonian constraints can be defined on $\cal 
H$ \cite{QSD}. However, since the classical Poisson algebra of 
constraints involves diff$(\sigma)$, it should come at no surprise that
the part of the quantum algebra that involves the Hamiltonian 
constraints does not manifestly mimic the classical algebra because 
in the quantum theory we can only define finite diffeomorphisms. In 
fact, there is a finite diffeomorphism analog for the commutator between 
diff$(\sigma)$ and the Hamiltonian constraints and that part of the 
algebra is realised without anomalies \cite{QSD}. However, the 
commutator between two Hamiltonian constraints classically is 
a linear combination, with phase space dependent coefficients, of 
elements of diff$(\sigma)$ and it is this commutator which is 
problematic in LQG.

The philosophy that has been adopted in \cite{QSD} is that the 
quantisation of the Hamiltonian constraints should be anomaly free in 
the sense that the (dual of the) commutator between two Hamiltonian 
constraints should annihilate the space of spatially diffeomorphism 
invariant states constructed in \cite{ALMMT}. This is indeed possible to 
achieve and one can show that this requires that the Hamiltonian 
constraints, which are densely defined on cylindrical functions, 
{\it necessarily change (enlarge)} the graph that underlies a given 
cylindrical function. This is also natural to happen because 
the natural regularisation of the constraint involves small loops that 
are attached to the vertices of a given graph which shrink towards the 
vertex 
as the regulator is removed. However, the shrinking process can be 
compensated for by a spatial diffeomorphism and since the limit is taken 
in an operator topology which involves spatially diffeomorphism 
invariant states, the loops actually do not completely shrink to the 
vertex. See \cite{QSD} or the second book in \cite{books} for details.

While the commutator of two Hamiltonian constraints then is anomaly free 
in the sense explained, in addition one would  
like to check that the classical limit of the commutator
between quantum Hamiltonian constraints is precisely the corresponding 
Poisson bracket between the classical constraints. Here again we are 
faced with an obstacle: For graph changing 
operators such as the Hamiltonian constraints it turns out to be 
extremely difficult to define coherent (or 
semiclassical) states. That is, states labelled by points in the 
classical phase space with respect to which the operator assumes an 
expection value which reproduces the value of the corresponding classical 
function at that point in phase space and with respect to which the 
(relative) fluctuations are small. The reason for why this happens is 
that the existing coherent states for LQG \cite{GCS} are defined over
a finite collection of finite graphs and these suppress very effectively 
the fluctuations of those degrees of freedom that are labelled by the 
given graph. However, the Hamiltonian constraints add degrees of freedom
to the state on which they act and the fluctuations of those are 
therefore no longer suppressed. 
Indeed, the semiclassical behaviour of the Hamiltonian 
constraints with respect to these coherent states is rather bad. \\
\\
Hence we see that the problem of investigating the classical limit of 
LQG and to verify the quantum algebra of constraints are very much 
interlinked:\\ 
1. Spatial diffeomorphism invariance enforces a weakly 
discontinuous representation of spatial diffeomorphisms.\\
2. Anomaly freeness in the presence of only finite diffeomorphisms 
enforces graph changing Hamiltonian constraints.\\
3. Graph changing Hamiltonians seem to prohibit appropriate 
semiclassical states.

\subsection{The Master Constraint Programme for LQG}
\label{s1.2}

The purpose of the Master Constraint Programme \cite{M1,M2} for 
LQG is to overcome those problems. The classical Master 
Constraint 
for a given 
(infinite) set of classical constraints is essentially the weighted sum 
of the 
squares of the individual constraints. The resulting Master Constraint 
carries the the same information about the reduced phase space as the 
original set of individual constraints. Since the infinite set of 
constraints was 
replaced by a single one, there are trivially no quantum anomalies no 
matter whether operators act in a graph changing or non -- graph 
changing fashion. 
However, whether or not the original quantum constraints that enter 
the construction of the Master Constraint are anomalous 
manifests itself in the spectrum of the Master Constraint \cite{Test}:
If the original algebra is anomalous then it is expected that zero is not 
contained in the 
spectrum of the Master Constraint. This can be cured by subtracting from 
the Master Constraint the minimum of the spectrum provided of course 
that it is finite and vanishes as $\hbar\to 0$ so that the modified 
constraint still has the same classical limit as the original one.
One then defines the physical Hilbert space as the (generalised) kernel 
of the Master Constraint, see the first reference of \cite{Test} for the 
mathematical details.

The Master Constraint for GR involves the weighted sum of squares of the 
Hamiltonian constraints such that the resulting expression is 
spatially diffeomorphism invariant. In 
\cite{M2} the Master Constraint has been quantised in two different 
ways: In the first version one used the 
graph changing operators defined in \cite{QSD}. Since the operator 
must be spatially diffeomorphism invariant, from the results of 
\cite{ALMMT} this operator must be defined directly on the 
spatially diffeomorphism invariant Hilbert space whose 
states are labelled by (generalised) knot classes.
The semiclassical analysis of this first operator is again difficult
because it changes knot classes and because so far no 
semiclassical spatially diffeomorphism invariant states have been defined in 
LQG. In the second version one used a non graph changing operator which 
therefore can be defined directly on the kinematical Hilbert space.
The Hamiltonian constraints that enter this operator would be anomalous, 
however, as we said, the Master Constraint does not care about this, 
moreover, the 
second Master Constraint is 
manifestly spatially diffeomorphism invariant.
The second operator therefore can in principle be analysed by existing 
semiclassical 
tools.

\subsection{Removing the Graph Dependence of Semiclassical States for 
LQG:\\ Algebraic Graphs}
\label{s1.3}

However, there is still one caveat: As already mentioned, the 
semiclassical tools for LQG developed
so far are based on pure states over single graphs or mixed states based 
on a certain class of graphs. None of these states involves all the graphs 
that are allowed in LQG and therefore those states cannot be semiclassical 
for all degrees of freedom of LQG. See e.g. the discussion in 
\cite{Complexifier}. One cannot sum over all graphs because the
sum is over uncountably many states, hence the state is not 
normalisable. Rather than taking an uncountable sum one could try to 
consider an uncountable tensor product which gives normalisable 
states \cite{ITP}. The problem here is that there is no such thing 
as a maximal graph in LQG of which all other graphs are subgraphs.\\
\\
Therefore, the existing semiclassical tools of LQG are heavily
{\it graph dependent}.\\
\\
It is at this point where we depart 
in a crucial way from LQG: We discard the notion of embedded graphs
and consider algebraic graphs instead. An algebraic graph is simply a 
labelling set consisting of abstract points (vertices) together with 
information how many abstract arrows (edges) point between points. There 
is no information about the knotting and braiding of those edges or about 
the location of the points. All that an algebraic graph knows about is the 
number of points and their oriented valence (that is, how many arrows 
point between different vertices with in or outgoing orientation).
Hence we lose information about topology and differential structure of 
the spatial manifold underlying LQG. We call the theory based 
on algebraic graphs Algebraic Quantum 
Gravity (AQG) in order to distinguish it from LQG by which it is 
inspired.

The point of introducing the notion of an algebraic graph is that it can 
be embedded in all possible ways into a given spatial manifold. Thus,
at least all embedded graphs with the same valence structure as the 
underlying algebraic graph can be obtained in this way and we will see 
that this is enough in order to do semiclassical physics because 
all physical (gauge invariant) operators, such as the Master Constraint, 
can be defined in an embedding 
independent (algebraic) fashion. One just has to lift the action of a 
given 
LQG operator on embedded graphs to the algebraic graph. What we have 
achieved by this is that our theory has lost its graph dependence, the
chosen algebraic graph is {\it fundamental} or maximal. 
It turns out that the algebraic graph necessarily must have a
countably infinite number of edges, see below.  

\subsection{The Extended Master Constraint}
\label{s1.4}

There is a problem with this idea which has to do with spatial 
diffeomorphism 
invariance: Since there is no $\sigma$ to begin with, there cannot be 
any Diff$(\sigma)$. Therefore, the natural action of the diffeomorphism 
group on the Hilbert space of LQG is not available in AQG. One could 
try to argue that the Hilbert space of AQG in some sense is already 
a space of spatially diffeomorphism invariant states. However, as shown 
in \cite{Corichi} this would make the physical Hilbert space of gauge 
invariant states too large.
Therefore one somehow must also perform a gauge reduction with respect 
to the spatial diffeomorphism constraints. Even if we forget about 
the fact that in AQG there is no Diff$(\sigma)$ and embed the 
algebraic graph into some $\sigma$ and thus consider a fixed embedded 
lattice, there are problems in defining lattice analogs of the 
generators 
of spatial diffeomorphisms, their algebra does not close for finite 
lattice length, see e.g. \cite{Loll}.   

It is at this point at which we invoke the {\it extended} Master 
Constraint introduced in \cite{M1}. 
The classical extended Master Constraint also involves the weighted sum 
of squares 
of the spatial diffeomorphism constraints such that the resulting 
expression is spatially diffeomorphism invariant. It can be quantised 
in LQG in a graph non changing and spatially diffeomorphism invariant 
fashion similar to the simple Master constraint. This may come as a 
surprise because the infinitesimal generator of spatial diffeomorphisms
cannot be defined in LQG. The solution of the puzzle is that the weight
function that enters the sum over squares becomes an operator which 
mildens the UV behaviour of the formally singular quantum generators
of spatial diffeomorphisms.
The point is now that this extended Master constraint also naturally 
lifts to algebraic graphs. This way we have achieved to implement 
also spatial diffeomorphism invariance on the algebraic level without 
running into anomalies.

Notice that many aspects of this idea to work at the embedding 
independent level had been spelled out already in \cite{STW}. However,
the programme could then not be pushed to its logical frontiers  
because it was unclear how to deal with spatial diffeomorphism 
invariance, that is, the (extended) Master Constraint 
programme was not yet developed. Also, there are certain operators in 
LQG such as the volume operator 
\cite{RSVol,ALVol,Volume} crucial for the quantum dynamics
which do carry embedding dependent 
information and therefore cannot be immediately lifted to the algebraic 
level. The way we deal with this here is that we choose a fixed 
algebraic graph once and for all and choose a {\it generic} 
embedding (this will be made precise later). We then 
lift the volume operator of LQG for those embeddings. This will mean 
that the semiclassical limit of this operator will come out right only
if the semiclassical states are defined using a generic embedding 
but again this turns out to be sufficient for semiclassical purposes.

\subsection{The Structure of AQG and Semiclassical States}
\label{s1.5}

As already mentioned, an algebraic graph does not contain any 
information 
about the braiding of its edges and is not embedded into any 3-manifold.
On such an algebraic graph one can define an abstract $^\ast-$ or 
$C^\ast-$algebra of elementary algebra elements out of which the Master 
Constraint is constructed as a composite operator. We use a specific 
representation of this algebra on a Hilbert space which is motivated from 
LQG and in this representation the Master constraint is a positive,
self-adjoint operator. In order to derive the classical limit of the 
theory we must give the following data: 1. a 3-manifold $\sigma$, 2.
initial data $m$ on $\sigma$ (equivalently: a point in the classical phase 
space, for gravity essentially a 3-metric and its extrinsic curvature) and 
3. an embedding of the algebraic graph (and a graph dual to it) into 
$\sigma$. Out of these data one can then construct a coherent state along 
the lines of \cite{GCS}.

In order that we can define a semiclassical limit 
for all $\sigma$ we must necessarily work with (countably) infinite 
algebraic graphs in order to be able to deal with asymptotically flat 
topologies.
If $\sigma$ is compact, the embedding of the algebraic graph will contain 
accumulation points but this is no obstacle for our formulation because we 
can leave all but finitely many of the edges (and dual faces) of the 
embedded graph unexcited thus effectively avoiding accumulation points.
This leads us naturally to von Neumann's Infinite Tensor Product 
(ITP) which was 
applied in the context of LQG in \cite{ITP}. Moreover, the ITP enables us 
to embed the algebraic graph as densely as we wish, thus making the 
semiclassical approximation as good as we like\footnote{This does not 
work for all operators of the theory but only for those which classically 
would come from volume integrals. Classical functions of this type 
separate the points of the classical phase space, see 
\cite{Complexifier} for a discussion.}. 

As an aside we should mention that while the (extended) Master 
Constraint can also defined in LQG in a non -- graph changing fashion,
such an operator is there rather ad hoc because one has to define it 
also on rather coarse graphs. On those graphs the expression for the 
operator proposed in \cite{M1} cannot be obtained by a regularisation 
process from the classical expression because the loops and edges 
involved might be ``large'. In contrast, in AQG there is a 
single graph 
to be considered and it is typically embedded in such a way that all 
loops and edges are small thus being close in appearance to the 
classical continuum expression.   

There is a crucial difference between the semiclassical states of LQG 
and of AQG: In both theories the coherent states are labelled by 
embedded graphs. However, in LQG these states are linear combinations of 
spin network functions\footnote{Spin network functions (SNWF) provide an 
orthonormal basis in LQG, in particular, SNWF's labelled by different 
graphs are orthogonal.} over the embedded graph with 
certain coefficients which carry the above data.  In AQG the 
coherent states are linear combinations of spin network functions
only if $\sigma$ is compact and even then these spin network functions 
are labelled by the unique 
abstract graph while the coefficients are labelled by the embedded 
graph. This tiny difference    
has e.g. the consequence that in LQG coherent states over different 
graphs are automatically orthogonal while in AQG this is not necessarily 
the case. 

Since we can accommodate any $\sigma$ in our formulation, AQG can 
presumably deal with 
topology change. Moreover, as was pointed out in \cite{ITP}, the 
non separable ITP is a direct sum of separable Hilbert spaces (sectors), 
some of 
which can be identified with excitations of our semiclassical states 
just discussed which could make contact with Fock spaces and low energy 
physics as sketched in \cite{ST}. 

Notice that in AQG, in contrast to 
the embedded graphs of LQG, the infinite algebraic graph is fixed. AQG
theories defined on different infinite algebraic graphs are unitarily 
equivalent if and only if there is a permutation of the vertices such that 
the algebraic graphs can be transformed into each other. Hence, in AQG the 
algebraic graph is a fundamental object. An interesting question is 
whether one could extend AQG in such a way as to accommodate all 
algebraic 
graphs. This seems neither necessary nor meaningful to us because one 
would 
need to relate the edges of different algebraic graphs to each other, 
however, without an embedding this is not possible\footnote{We could 
declare the Hilbert spaces labelled by different algebraic graphs as 
orthogonal to each other where different means that there is no 
permutation transformation between the corresponding adjacency matrices,
see section \ref{s2}.
The elementary operators of the theory would then also be labelled by 
the 
algebraic graph in addition to edges and vertices and one would embed 
different algebraic graphs in such a way that they are disjoint in order 
to be consistent with LQG where such states would be orthogonal.
However, there seems to be no physical justification for such a choice at 
present.}. 

Therefore in this 
paper and the companion paper we will focus on cubic algebraic graphs
(all vertices have valence six) which will simplify our calculations 
and turns out to be sufficient in order to semiclassical calculations.

Notice that no continuum limit has to be performed on 
the algebraic graph. None of the operators of the theory 
depends on a lattice length. This is not possible because the 
theory is manifestly background independent.
There are no scales to be sent to zero, everything is UV 
finite. The precision with which the semiclassical limit is reached 
depends on the choice of the embedding (its ``fineness'' with respect 
to the background metric to be approximated) which is a feature of the 
state but 
the fundamental quantum algebra does not know about this. This is in 
strong contrast to, say lattice QCD on Minkowski space, where the 
Hamiltonian depends explicitly on the latice length. One could 
interpret 
AQG as saying that lattice calculations are correct and that the 
lattice is actually fundamental if it is thought of as the 
concrete embedding of the algebraic graph. Lattice refinements are then 
to be thought of as different choices of embeddings of the fundamental 
algebraic graph. This also sheds new light on Wilson's notion 
of the 
renormalisation group.

\subsection{The Semiclassical Limit}
\label{s1.6}

With this set -- up, in 
two companion papers \cite{II,III} we will establish that 
the semiclassical limit of the extended Master Constraint is correct.
More precisely, in \cite{II} we carry out an exact computation using a
simplification which consists in replacing the non -- Abelean
group $SU(2)$ by the Abelean group $U(1)^3$. This computation reproduces 
the classical $U(1)^3$ analog of the Master constraint to zeroth order 
in $\hbar$. The point of this approximation  
is that the $U(1)^3$ analog of the volume operator, which enters the 
Master constraint in a pivotal way, is analytically diagonisable. This
is not the case for $SU(2)$ and prohibits exact semiclassical 
calculations. In \cite{III} we develop semiclassical perturbation theory 
for AQG and LQG {\it with error control} which allows to 
analytically calculate coherent state
matrix elements of positive fractional powers the $SU(2)$ volume 
operator up to any order in 
$\hbar$. The resulting semiclassical $SU(2)$ calculation is then 
exactly analogous to the $U(1)^3$ and reproduces the same classical 
limit as follows from the results of \cite{GCS}. Hence \cite{II,III} 
together imply that the infinitesimal gauge generators of AQG have 
the correct classical limit. This is what is so far missing in LQG. 

The coherent states chosen maybe further improved for instance by 
statistical averaging over a certain class of embedded graphs as to 
produce a density matrix, see e.g. 
\cite{Bombelli}. Notice that here again there is a crucial difference 
between LQG and AQG: In LQG the statistical average of coherent 
states, which are linear combinations of spin network states, 
affects both the 
spin network states and their coefficients. In AQG it affects only the 
coefficients. Let $\Gamma$ be some uncountable set of graphs embedded 
into some 
$\sigma$ and let $\mu$ be a 
probability measure on $\Gamma$. Let $P_{\psi_\gamma}$ be the 
projection onto the coherent state $\psi_\gamma$ 
and consider the object $\rho:=\int_\Gamma\; d\mu(\gamma)\; 
P_{\psi_\gamma}$. Then it is not difficult to see \cite{Complexifier} 
that in LQG this operator is the zero operator while 
in AQG this operator is trace class with unit trace. Even if we formally
interchange the integral over $\Gamma$ with taking the trace there are 
still qualitative differences, for generic operators $A$ in LQG 
which 
admit an embedding independent lift to AQG such as the total volume of a 
compact manifold $\sigma$ between the corresponding values of  
Tr$(A\rho)$.   

\subsection{Summary of Differences between AQG and LQG}
\label{s1.7} 

For the benefit of the reader we summarise the most important 
conceptual differences and similarities between AQG and LQG in the 
subsequent table.\\
\begin{table}
\center
\begin{tabular}{|l|c|l|}
\hline
 {\bf Object} & {\bf LQG} &{\bf AQG}\\ \hline\hline
Topology & must be provided & absent\\ \hline
Differentiable structure &  must be provided & absent\\ \hline
Hilbert space & ${\cal H}_{LQG}:={\cal H}_{AIL}$ & ${\cal H}_{AQG}:={\cal H}^{\otimes}$\\
\hline
Separability & non -- separable & non -- separable\\ \hline
graphs & embedded & algebraic (combinatorical)\\ \hline
$\#$ graphs & uncountably infinite & one \\ \hline
Structure of graphs & finite & countably infinite\\ \hline
Generating set of $^\ast-$algebra $\mathfrak{A}$ &
uncountably infinite & countably infinite\\ \hline 
\end{tabular}
\caption{Summary of the major differences in the mathematical structure of Loop Quantum Gravity (LQG) and Algebraic Quantum Gravity (AQG).}
\end{table}

Notice that the reason for the Hilbert spaces to be non separable 
is very different in the two cases: For LQG it is due to the fact the 
set $\Gamma$ of all finite embedded graphs is uncountable.
For AQG it is due to the fact that the ITP of a countable 
number of Hilbert spaces of which at least countably infinite many are 
at least two dimensional is not separable. Also the two Hilbert spaces 
of LQG and AQG are not directly related to each other. The only thing 
one can say is the following: Given an algebraic graph, a manifold 
$\sigma$ and an embedding $X$ we can consider the set $\Gamma^X_\alpha$
of all finite subgraphs of $X(\alpha)$. Consider the closed linear span
${\cal H}^X_\alpha$ of spin network states over elements of 
$\Gamma^X_\alpha$. Then ${\cal H}^X_\alpha\subset {\cal H}_{{\rm LQG}}$
for all $X$. On the other hand, for all $X$ the spaces ${\cal 
H}^X_\alpha$ are isomorphic to the sector of ${\cal H}_{{\rm AQG}}$
which is the closed linear span of finite excitations of the vector 
$\otimes_e {\rm\bf 1}$ where ${\rm\bf 1}$ is the constant function equal 
to unity.
 
Notice that in LQG one needs all 
graphs because the algebra of elementary operators contains the 
holonomies along all possible paths and those are obtained from a 
fixed given path through the 
natural action of the diffeomorphism group. In AQG the action of the 
infinitesimal
diffeomorphisms preserves the algebraic graph and so there is no need 
to take all algebraic graphs into account. This is different from what 
was done in \cite{ITP} where one worked in an embedding dependent 
context and considered ITP Hilbert spaces over all possible countably 
infinite embedded graphs.

\subsection{Organisation of the Article}
\label{s1.8}

In section two we introduce the concept of an algebraic graph and define
the abstract $^\ast-$algebra labelled by it. 
For an arbitrary algebraic graph we introduce the extended Master 
constraint using the notion of a minimal loop.

In section three we review the framework of coherent states as developed 
in \cite{GCS} as well as elements of the infinite tensor product 
construction of \cite{ITP} and lift it to the algebraic level.

In section four we present the result of the calculation of our companion 
papers \cite{II,III} which establishes the correctness of the classical 
limit of the 
Master Constraint on cubic algebraic graphs.

In section five we forecast the tasks that can be addressed using the 
new 
AQG framework. In particular we have in mind applications in quantum 
cosmology and the contact with the physics of the standard model. In order 
to do so one has to deal with the question in which sense one can perform 
trustable computations without solving the 
theory\footnote{That is, the construction of 1.
physical states annihilated by the Master Constraint, 2. Operators 
commuting with it as well as 3. a definition of the quantum dynamics in 
terms 
of physical Hamiltonians.}. We present a possible scheme, elements of 
which were proposed in \cite{Cosmo}, which could be called {\it Quantum 
Gauge Fixing}.
 
In section six we sketch how ideas from AQG might help to solve two 
important problems for the spin foam approach to LQG, namely 1. to make 
contact 
with the canonical programme which as we prove in \cite{II,III} does 
have
contact to the classical theory and 2. to get rid of the triangulation 
dependence of spin foam models.

Finally, in section seven we summarise and list some interesting open 
problems. 

\section{Algebraic Quantum Gravity}
\label{s2}

As appropriate for a top to bottom approach we introduce the basic 
ingredients of AQG axiomatically. In a second step we show how to 
extract physics from the mathematical notions and in particular reveal 
the connection with LQG. The latter is the subject of sections \ref{s3} 
and \ref{s4}.

\subsection{Algebraic Graphs}
\label{s2.1}

Comprehensive monographs on algebraic graphs are listed in \cite{Hofer}.
Here we just summarise what is needed at this point for our purposes.\\
\\
An (oriented) algebraic graph with $N$ vertices can be defined in 
terms of its adjacency matrix. This is an $N\times N$ matrix $\alpha$ 
whose entries 
$\alpha_{IJ}$ take non negative integer values $n$ where $n$ denotes the 
number
of edges that start in vertex $i$ and end in vertex $j$. Notice that 
$\alpha_{IJ},\;\alpha_{JI}$ are not related to each other and that      
$e_{IJ}:=\alpha_{IJ}+\alpha_{JI}$ is the total number of edges that 
connect vertices $I,J$. The valence of $I$ is given by $v_I=\sum_J 
e_{IJ}$. We also use the symbols $V(\alpha),\;E(\alpha)$
to denote the set of vertices and edges respectively and 
$b(e),\;f(e)$ to denote the vertex at which $e$ begins or finishes 
respectively. 
 
We are only interested in oriented algebraic graphs but for completeness 
we mention that for unoriented algebraic graphs the adjacency matrix is 
symmetric, its entries $\alpha_{IJ}$ being the total number of edges 
connecting vertices $I,J$ and $v_I=\sum_J \alpha_{IJ}$ is the valence of 
vertex $I$. We will be interested in $N=\aleph$, i.e. graphs where the 
number of edges has countably infinite cardinality but where the valence 
of each 
vertex is bounded by a small number of order unity, typically by $2D$ for 
cubic algebraic graphs or $D+1$ for simplicial algebraic graphs which we 
wish to embed into a $D-$dimensional manifold. This is necessary in order 
that the semiclassical limit of the theory is reached for arbitrary non 
compact topologies $\sigma$.

There is no information contained in the adjacency matrix which 
tells us how the various edges are braided. Also no information is 
available whether the edges are smooth, or $n-$times differentiable, 
whether the tangents of two edges adjacent at a vertex intersect there at 
a non -- vanishing angle etc. In particular, cubic 
algebraic graphs ``with defects'', i.e. those obtained by deleting $D-1$ 
edges adjacent 
at each vertex (the degrees of freedom on the deleted edges are then not 
excited) can be considered as algebraic simplicial graphs, hence the 
simplicial case is contained in the cubic one. This might be of some 
importance because it is easy to generate random, simplicial, embedded 
graphs by the Dirichlet -- Voronoi procedure \cite{Bombelli} which 
improves the semiclassical properties of coherent states or density 
matrices constructed from them \cite{Complexifier}.

\subsection{Quantum Kinematics}
\label{s2.2}

\subsubsection{Gauge Field and Gravitational Sector}
\label{s2.2.1}

Given an algebraic graph 
$\alpha$ we associate with each of its (distinguishable) edges $e$ an
element $A(e)$ of a compact, connected, semisimple Lie group $G$ and an 
element $E(e)$ of\footnote{For LQG practioners we stress that the 
notation $E(e)$ is no misprint: $E$ is just labelled by the edges of the 
algebraic graph, surfaces will come in only when we consider 
semiclassical states.} its Lie algebra Lie$(G)$.

These are subject to the algebraic relations 
\ba \label{2.1}
{[}A(e),A(e')] &=& 0\\
{[}E_j(e),A(e')] &=& i\hbar Q^2 \delta_{e,e'} \tau_j/2 A(e)
\nonumber\\
{[}E_j(e),E_k(e')] &=& -i\hbar Q^2 \delta_{e,e'} f_{jkl} E_l(e')
\nonumber
\ea
Furthermore, the following $^\ast-$relations hold
\be \label{2.1a}
A(e)^\ast=[A(e)^{-1}]^T,\;\;E_j(e)^\ast=E_j(e)
\ee
We will denote the resulting\footnote{We have set the gravitational 
Immirzi parameter to unity, otherwise rescale $Q$ appropriately.} 
$^\ast-$algebra by $\mathfrak{A}$.
Here $Q^2$ plays the role of the coupling constant of the gauge 
theory\footnote{In particular $Q^2=8\pi G_{{\rm Newton}}$ for gravity. 
If needed we write $Q_{{\rm GR}}$ for gravity and $Q_{{\rm YM}}$ for 
the gauge field sector.} in 
question and $\tau_j,\;j=1,..,\dim(G)$ are generators of the Lie algebra 
of $G$ which we take to be anti -- Hermitean and trace -- free for 
convenience since any compact, semisimple 
Lie group can be realised as a subgroup of some $SU(N)$. These satisfy
$[\tau_j,\tau_k]=f_{jkl} \tau_l$ where the structure constants $f_{jkl}$ 
are totally skew and we normalise according to Tr$(\tau_j 
\tau_k)=-\frac{1}{2}\delta_{jk}$. Also $E_j(e):=-2{\rm Tr}(\tau_j 
E(e))$. Obviously, (\ref{2.1}) takes the form 
of a direct sum of $^\ast-$algebras, one for each edge $e$, each of which 
can be considered as the quantisation of the cotangent bundle $T^\ast(G)$.

A natural representation of the algebra $\mathfrak{A}$ in (\ref{2.1}) is 
the infinite tensor product (ITP) Hilbert space ${\cal 
H}^\otimes:=\otimes_e {\cal H}_e$ where ${\cal H}_e\cong L_2(G,d\mu_H)$ 
and $\mu_H$ is the Haar measure on $G$. Other representations are 
conceivable but this representation is natural if we want to match the 
uniqueness result \cite{LOST,Fleischhack} of LQG valid for any 
(semianalytic) 3-manifold. For a review of the ITP
and the associated von Neumann algebras connected with it see e.g. 
\cite{ITP}. We just collect the necessary notions here.

The ITP Hilbert space is closure of the finite linear span of vectors of 
the form $\otimes_f:=\otimes_e f_e$ where $f_e\in {\cal H}_e$. The inner
product between these vectors is given by
\be \label{2.2}
<\otimes_f,\otimes_{f'}>:=\prod_e \;<f_e,f'_e>_{{\cal H}_e}
\ee
The infinite product $\prod_e z_e$ of complex numbers $z_e=|z_e| 
e^{i\phi_e}$ is 
defined by $\prod_e z_e:=[\prod_e |z_e|]\; e^{i\sum_e \phi_e},\;\phi_e\in 
[-\pi,\pi)$ provided 
that both of $\sum_e ||z_e|-1|$ and $\sum_e |\phi_e|$ converge, in which 
case we also say that $\prod_e z_e$ is convergent. 
Otherwise we set $\prod_e z_e=0$. One can show that for $z=\prod_e 
z_e\not=0$ we can find for any $\delta>0$ a finite subset 
$E_\delta(\alpha)$ of the set $E(\alpha)$ of edges of $\alpha$ such 
that $|z-\prod_{e\in E_\delta(\alpha)} z_e|<\delta$ for all 
$E_\delta(\alpha)\subset 
E\subset E(\alpha)$. Obviously we consider only elements such that 
$||\otimes_f||\not=0$. 
  
Two vectors $\otimes_f,\; \otimes_{f'}$ are said to be strongly equivalent
if and only if $\sum_e |<f_e,f'_e>_{{\cal H}_e}-1|$ converges. We denote 
by $[f]$ the strong equivalence class of $f$. It follows that 
$<\otimes_f,\otimes_{f'}>=0$ if either $[f]\not=[f']$ or 
$[f]=[f']$ and $<f_e,f'_e>=0$ for at least one $e$.   

We say that $\prod_e z_e$ is quasi convergent if $\prod_e |z_e|$ 
converges. If we set $(z\cdot f)_e:=z_e f_e$ then $\otimes_{z\cdot f}=
(\prod_e z_e)\;\otimes_f$ fails to hold if $\prod_e z_e$ is not 
convergent. We say that $f,f'$ are weakly equivalent provided that there 
exists $z$ such that $[z\cdot f]=[f']$. This is equivalent to the 
convergence of $\sum_e |\;\;|<f_e,f'_e>|-1|$. We denote by $(f)$ the 
weak 
equivalence class of $f$. Obviously, strong equivalence implies weak 
equivalence. One can show that the closure of the span of all vectors in 
the same strong equivalence class $[f]$, denoted by ${\cal 
H}^\otimes_{[f]}$  is separable, consisting of the 
completion of the finite linear span of the vectors of the form 
$\otimes_{f'}$ where $f'_e=f_e$ for all but finitely many $e$. The ITP 
Hilbert space ${\cal H}^\otimes$ is the direct sum of the ${\cal 
H}^\otimes_{[f]}$. Let also ${\cal H}^\otimes_{(f)}$ be the closure of the 
finite linear span of the $\otimes_{f'}$ with $(f')=(f)$. Then the strong 
equivalence subspaces of ${\cal H}^\otimes_{(f)}$ are unitarily 
equivalent, the corresponding unitary operators being of the form $U_z 
\otimes_f:=\otimes_{z\cdot f}$ with $\prod_e z_e$ quasi convergent.  

Our basic operators act in the obvious way as  
\ba \label{2.3}
A(e)\otimes_f &:=& [A(e) f_e]\; \otimes \; [\otimes_{e'\not=e} f_{e'}]     
\\
E_j(e)\otimes_f &:=& [E_j(e) f_e]\; \otimes \; [\otimes_{e'\not=e} f_{e'}]     
\nonumber
\ea
where $[A(e) f_e](h):=h f_e(h)$ and 
$[E_j(e) f_e](h):=i\hbar Q^2 [\frac{d}{dt}]_{|t=0} f_e(e^{t\tau_j/2}h)$.
It is not difficult to show that this makes $A(e)$ a unitary matrix valued
(in particular bounded) multiplication operator and $E_j(e)$ an 
essentially self -- adjoint derivation operator. 
The relations (\ref{2.3}) define them densely on ${\cal H}^\otimes$. 
This concludes the definition of the quantum kinematics. 

\subsubsection{Fermionic Sector}
\label{s2.2.2}

Given an algebraic graph $\alpha$ we associate with each vertex $v\in 
V(\alpha)$ Grassmann -- valued variables $\theta_M(v),\bar{\theta}_M(v)$
where $M$ is a compound index $M\equiv (m,I)$ where $m=\pm 1/2$ is a 
Weyl 
spinor index and $I=1,..,d$ where $d$ is the dimension of the 
defining representation of the Yang -- Mills group $G$. These are subject
to the anti -- commutation relations
\be \label{2.3a}
[\theta_M(v),\theta_N(v')]_+=[\bar{\theta}_M(v),\bar{\theta}_N(v')]_+=0,\;\;
[\theta_M(v),\bar{\theta}_N(v')]_+=Q_{{\rm F}}^2 \hbar \delta_{MN} 
\delta_{v,v'}
\ee
as well as the $^\ast-$relations
\be \label{2.3b}
[\theta_M(v)]^\ast=\bar{\theta}_M(v)
\ee
Here $\hbar Q^2$ is dimensionfree if we take $\theta$ to be
dimensionless\footnote{The $\theta$ are related to the usual fermionic degrees
of freedom of dimension cm$^{-3/2}$ by a canonical transformation which takes
care of the dimensionalities, see below.}.
We consider just one fermion species and only one helicity\footnote{By the 
canonical
transformation (it preserves anti -- Poisson brackets) $\theta_m\mapsto
\bar{\theta}_m$ one can switch between left and right handed descriptions.}.
Again we will denote the resulting $^\ast-$algebra by $\mathfrak{A}$. 
A natural representation thereof is again by an infinite tensor product:
For each $v$ we consider the $2^{2d}$ dimensional Hilbert space of 
``holomorphic'' functions\footnote{Notice that $\theta,\bar{\theta}$ are
classically anticommuting Grassman numbers but that in quantum theory
classical identities such as the nilpotency $[\theta\bar{\theta}]^2=0$ 
no
longer holds.}   
$f_v(\theta)=\sum_{k=0}^{2d} \sum_{1\le M_1<..<M_{2d}} f_v^{M_1..M_k} 
\theta_{M_1}(v) .. 
\theta_{M_k}(v)$
where the complex valued coefficients are totally skew. Set 
for one single Grassman degree of freedom $d\mu(\theta)=
d\theta d\bar{\theta} (1+\bar{\theta}\theta/(\hbar Q_{{\rm F}}^2))$ and 
define the 
usual 
Berezin ``integral'' over superspace (better: linear functional) $\int 
d\theta 1=0,\;\int d\theta \theta 
=1$. We now consider the infinite tensor product 
${\cal H}^\otimes:=\otimes_{v\in V(\alpha)} {\cal H}_v$ where 
${\cal H}_v=L_2(d\mu_v),\;d\mu_v(\theta)=\prod_M d\mu(\theta_M(v))$ which 
is a representation space of $\mathfrak{A}$ via 
$(\theta_M(v) f_v)(\theta):=\theta_M(v) f_v(\theta)$ and 
$(\bar{\theta}_M(v) f_v)(\theta):=\hbar Q_{{\rm F}}^2 
\partial/\partial\theta_M(v) 
f_v(\theta)$ (left derivative). 

All remarks about the infinite tensor product from the last subsection 
apply, just that the label set has switched from edges to vertices.

\subsubsection{Higgs Sector}
\label{s2.2.3}

Given an algebraic graph $\alpha$ we associate to each vertex $v\in 
V(\alpha)$ Lie$(G)$ valued (if the Higgs transforms in the adjoint 
representation) or vector valued (if the Higgs transforms in the defining 
representation of $G$) fields $\phi_j(v),\pi_j(v)$ subject to the 
algebraic relations
\be \label{2.3c}
[\phi_j(v),\phi_k(v')]=[\pi_j(v),\pi_k(v')]=0,\;\;
[\pi_j(v),\phi_k(v')]=i\hbar Q_{{\rm H}}^2 \delta_{jk} \delta_{v,v'}
\ee
and the $^\ast-$relations
\be \label{2.3d}
\phi_j(v)^\ast=\phi_j(v),\;\;\pi_j(v)^\ast=\pi_j(v)
\ee
if the Higgs is Lie$(G)$ valued. If it transforms in the defining 
representation so that it is complex valued then we split the Higgs into 
real and imaginary part and impose (\ref{2.3d}) on those.

Again an infinite tensor product provides a representation of this 
$^\ast-$algebra $\mathfrak{A}$. Consider the probability measure on $\Rl$ 
given by $d\mu(x)=e^{x^2/2}dx/\sqrt{2\pi}$ and $d\mu_v(\phi):=\prod_j 
d\mu(\phi_j(v))$. Let ${\cal H}_v=L_2(d\mu_v)$ and ${\cal H}^\otimes
=\otimes_v {\cal H}_v$. We consider functions of the form $f_v(\phi)\equiv
f_v(\{\phi_j(v)\}_j)$ which depend only on the $\phi_j(v)$. Then
$[\phi_I(v) f_v](\phi):=\phi_I(v) f_v(\phi)$ and 
$[\pi_I(v) f_v](\phi):=i\hbar[\partial/\partial\phi_I(v)-\phi_I(v)/2] 
f_v(\phi)$ provide a representation of $\mathfrak{A}$ on ${\cal 
H}^\otimes$.

\subsection{Quantum Dynamics}
\label{s2.3}

We turn now to the quantum dynamics. Pivotal for everything to come 
is the volume operator:\\
Given a vertex $v$ of the algebraic graph we set
\be \label{2.4}
V_v:=\ell_P^3\sqrt{|\frac{1}{48} \sum_{e_1\cap e_2\cap e_3=v}
\epsilon_v(e_1,e_2,e_3) \epsilon^{ijk} E_i(e_1) E_j(e_2) E_k(e_3)|}
\ee
where the sum is over all triples of mutually distinct edges $e_1,e_2,e_3$ 
incident 
at $v$. The totally skew function 
$(e_1,e_2,e_3)\mapsto \epsilon_v(e_1,e_2,e_3)$ takes values 
$\pm 1,0$ and
will be chosen according to the algebraic graph in question in such a
way that it matches the embedding dependent volume operator of LQG
\cite{ALVol} when embedding the algebraic graph in a generic\footnote{The 
possible embeddings of an algebraic 
graphs fall into diffeomorphism equivalence classes. An embedding is 
called generic if a random embedding results with non vanishing 
probability in an embedded graph of the same equivalence class. If 
there is more than one possibility then we must pick one. For our
cubic graph to be considered later we consider half -- generic 
embeddings in the sense that there is a neighbourhood of each vertex 
and a coordinate system in which the graph looks like the three 
coordinate axes in $\Rl^3$.}
way. The functions $\epsilon_v(e_1,e_2,e_3)$ are then chosen once and 
for all, they are embedding independent. Notice 
that the embedding independent operator 
\cite{RSVol} has been ruled out as inconsistent in a recent analysis
\cite{GT1,GT2}. In formula (\ref{2.4}) we have assumed that all edges 
are outgoing from $v$. If $e$ is ingoing at $v$, then replace $E_j(e)$ by
$-{\rm Ad}_{jk}(h_e) E_k(e)$ where $h\tau_j h^{-1}=:{\rm Ad}_{jk}(h) 
\tau_k$ defines the adjoint representation of $G$ on Lie$(G)$. 

We will also need the total volume given by $V=\sum_{v\in V(\alpha)}
V_v$. Finally we need the crucial operators 
\be \label{2.10}
Q^{(r)}_v=\frac{1}{T_v}\sum_{e_1\cap e_2\cap e_3=v} 
\epsilon_v(e_1,e_2,e_3) 
{\rm Tr}(
(A(e_1) [A(e_1)^{-1},V_v^r]))
(A(e_2) [A(e_2)^{-1},V_v^r]))
(A(e_3) [A(e_3)^{-1},V_v^r]))
) \ee
where $T_v$ is the number of unordered triples of mutually distinct edges 
incident at $v$ and $r$ is any real number. They will be needed in order 
to ensure the correct density weight of the various expressions in the 
classical limit of the Master Constraint.\\
\\
We now consider the following composite operators the classical limit of 
which are half densities.

\subsubsection{Gravitational Sector}
\label{s2.3.1}
\begin{itemize}
\item[A.1a] {\it Gravitational Gauss Constraint}\\
For any $v\in V(\alpha)$ we set
\be \label{2.5}
G^{GR}_j(v):=Q_v^{(1/2)} [\sum_{b(e)=v} E_j(v)-\sum_{f(e)=v} {\rm 
Ad}_{jk}(A(e)) E_k(v)]
\ee 
where $j,k=1,2,3$ for $G=SU(2)$.
\item[B.1] {\it Spatial Diffeomorphism Constraint}\\
Given a vertex $v$ of the algebraic graph $\alpha$ and two edges $e,e'$ 
incident
at and outgoing from $v$, a loop $\beta_{v,e,e'}$ within $\alpha$ starting 
at $v$ along
$e$ and ending at $v$ along $(e')^{-1}$ is said to be minimal \cite{M1}
provided that there is no other loop within $\alpha$ satisfying the same 
restrictions with fewer edges traversed. We denote by $L(v,e,e')$ the 
set of minimal loops with the data indicated.

For any $v\in V(\alpha)$ we set
\be \label{2.6}
D^{GR}_j(v):=\frac{1}{T_v}\sum_{e_1\cap e_2\cap e_3=v} 
\frac{\epsilon_v(e_1,e_2,e_3)}{|L(v,e_1,e_2)|} \sum_{\beta 
\in L(v,e_1,e_2)}
{\rm Tr}(\tau_j [A(\beta)-A(\beta)^{-1}] A(e_3) [A(e_3)^{-1},\sqrt{V_v}])
\ee 
where the sum is over unordered triples of mutually distinct edges 
adjacent to $v$ 
and where again we assumed for convenience that all edges are outgoing 
from $v$. The quantity $T_v:=|\{e_1\cap e_2\cap 
e_3=v;\;|\epsilon_v(e_1,e_2,e_3)|=1\}|$ is the number of contributing 
triples. 
\item[C.1a] {\it Euclidean Hamiltonian Constraint}\\
For any $v\in V(\alpha)$ and any $0<r\in \Ql$ we set
\be \label{2.7}
H^{(r)}_E(v):=\frac{1}{T_v}\sum_{e_1\cap e_2\cap e_3=v} 
\frac{\epsilon_v(e_1,e_2,e_3)}{|L(v,e_1,e_2)|} 
\sum_{\beta 
\in L(v,e,e')}
{\rm Tr}([A(\beta)-A(\beta)^{-1}] A(e_3) [A(e_3)^{-1},(V_v)^r])
\ee 
where the conventions are the same as above. This constraint is just
an auxiliary construction which we need in order to define various other 
quanties, it has no physical meaning in our manifestly Lorentzian theory.
\item[C.1b] {\it (Lorentzian) Hamiltonian Constraint}\\
For any $v\in V(\alpha)$ we set
\ba \label{2.8}
&& H^{GR}(v)-H^{(1/2)}_E(v) := \frac{1}{T_v}\sum_{e_1\cap e_2\cap 
e_3=v}
\epsilon_v(e_1,e_2,e_3)  
\times \\ && \times
{\rm Tr}( 
(A(e_1) [A(e_1)^{-1},[H^{(1)}_E,V]]))
(A(e_2) [A(e_2)^{-1},[H^{(1)}_E,V]]))
(A(e_3) [A(e_3)^{-1},\sqrt{V_v}]))
\nonumber
\ea
where the conventions are the same as above and $H_E^{(1)}:=\sum_v 
H_E^{(1)}(v), \; V:=\sum_v V_v$.
\end{itemize}

\subsubsection{Yang -- Mills Sector}
\label{s2.3.2}

\begin{itemize}
\item[A.2b] {\it Yang -- Mills Gauss Constraint}\\
For any $v\in V(\alpha)$ we set
\be \label{2.8a}
\underline{G}^{YM}_J(v):=
Q_v^{(1/2)} [\sum_{b(e)=v} \underline{E}_J(v)-\sum_{f(e)=v} 
{\rm 
Ad}_{JK}(\underline{A}(e)) \underline{E}_K(v)]
\ee 
where $J,K=1,..,\dim(G)$ for $G$ and we use underlined symbols to 
distinguish gravitational and Yang -- Mills quantities.
\item[B.2] {\it Spatial Diffeomorphism Constraint}\\
For any $v\in V(\alpha)$ we set
\be \label{2.8b}
D^{YM}_j(v):=Q_v^{(1/6)}\frac{1}{P_v}\sum_{e_1\cap e_2=v} 
\frac{1}{|L(v,e_1,e_2)|} \sum_{\beta 
\in L(v,e_1,e_2)}
{\rm Tr}([\underline{A}(\beta)-A(\beta)^{-1}]\underline{E}(e_1))
E_j(e_2) 
\ee 
where the sum is over all pairs of distinct edges adjacent to $v$ and 
$P_v$ is their number.
\item[C.2] {\it Hamiltonian Constraint}\\
For any $v\in V(\alpha)$ we set 
\ba \label{2.8c}
H^{YM}(v) &=& \frac{1}{2Q^2}
\frac{1}{P'_v}\sum_{e_1\cap e_2=v} 
[{\rm Tr}(\tau_j A(e_1) [A(e_1)^{-1},V_v^{1/4}]) 
\underline{E}_J(e_1)]^\dagger
[{\rm Tr}(\tau_j A(e_2) [A(e_2)^{-1},V_v^{1/4}]) 
\underline{E}_J(e_2)]
\nonumber\\
&& +
\frac{1}{T_v}\sum_{e_1\cap e_3\cap e_4=v} 
\frac{1}{T_v}\sum_{e_2\cap e_5\cap e_6=v} 
\frac{\epsilon_v(e_1,e_3,e_4)}{|L(v,e_3,e_4)|} 
\frac{\epsilon_v(e_2,e_5,e_6)}{|L(v,e_5,e_6)|} 
\sum_{\beta \in L(v,e_3,e_4)}
\sum_{\beta' \in L(v,e_5,e_6)}
\times\nonumber\\
&& \times
[{\rm Tr}(\tau_j A(e_1) [A(e_1)^{-1},V_v^{1/4}]) 
{\rm Tr}(\underline{\tau}_J \underline{A}(\beta))]\;
[{\rm Tr}(\tau_j A(e_2) [A(e_2)^{-1},V_v^{1/4}]) 
{\rm Tr}(\underline{\tau}_J \underline{A}(\beta'))]
\nonumber\\
&&
\ea
Here in the electric term we sum over all pairs of edges incident at $v$
and $P'_v$ is their number.
\end{itemize}

\subsubsection{Fermionic Sector}
\label{s2.3.3}

\begin{itemize}
\item[A.3a] {\it Gravitational Gauss Constraint}\\
For any $v\in V(\alpha)$ we define
\be \label{2.8d}
G^F_j(v)=Q_v^{(1/2)}\sum_I\bar{\theta}_{mI}(v) (\tau_j)_{mn} 
\theta_{nI}(v)
\ee
\item[A.3b] {\it Yang -- Mills Gauss Constraint}\\
For any $v\in V(\alpha)$ we define
\be \label{2.8e}
\underline{G}^F_J(v)=Q_v^{(1/2)}\sum_m \bar{\theta}_{mI}(v) 
(\underline{\tau}_J)_{IK} 
\theta_{mK}(v)
\ee
\item[B.3] {\it Spatial Diffeomorphism Constraint}\\
For any $v\in V(\alpha)$ we define
\be \label{2.8f}
D^F_j(v):=\frac{i}{2}\sum_{b(e)=v} Q_v^{(1/6)} E_j(e)
[\bar{\theta}_{mJ}(v) [A(e)]_{mn}
[\underline{A}(e)]_{JK} \theta_{nK}(f(e))-h.c.]
\ee
where as usual h.c. denotes the adjoint, with respect to our chosen 
representation, of the expression in the bracket and the sum is over all 
edges adjacent to $v$ which are outgoing from there\footnote{This 
corresponds to the forward lattice derivative. One can also add a term 
involving the incoming edges adjacent to $v$ corresponding to the backward 
lattice derivative.}.  
\item[C.3] {\it Hamiltonian Constraint}\\
For any $v\in V(\alpha)$ we define 
\ba \label{2.8g}
H^F(v) &=& 
\sum_{b(e)=v} Q_v^{(1/2)} E_j(e) 
\times \nonumber\\
&&\{ 
[Q_{f(e)}^{(1/2)}]^2 (A(e))_{jk}
\bar{\theta}_{mJ}(f(e))(\tau_j)_{mn} \theta_{nJ}(f(e))
-[Q_v^{(1/2)}]^2
\bar{\theta}_{mJ}(v)(\tau_j)_{mn} \theta_{nJ}(v)
\nonumber\\
&& +i [Q_v^{(1/2)}]^2  
[\bar{\theta}_{mJ}(v) [A(e)]_{pn}
[\underline{A}(e)]_{JK} (\tau_j)_{mp} \theta_{nK}(f(e))-h.c.]
\nonumber\\
&& - [Q_v^{(1/2)}]^2 {\rm Tr}(\tau_j 
A(e)[A(e)^{-1},[H_E^{(1)}(1),V]])\;
\bar{\theta}_{mJ}(v) \theta_{mJ}(v)
\}
\ea
Here $(A(e))_{jk}$ denotes the matrix elements of the holonomy in the 
spin one representation.
\end{itemize}

\subsubsection{Higgs sector}
\label{s2.3.4}

\begin{itemize}
\item[A.4b] {\it Yang -- Mills Gauss Constraint}\\
For any $v\in V(\alpha)$ we define
\be \label{2.8h}
\underline{G}^H_J(v)= Q_v^{(1/2)} \pi_K(v) (\underline{\tau}_J)_{KL} 
\phi_L(v)
\ee
\item[B.4] {\it Spatial Diffeomorphism Constraint}\\
For any $v\in V(\alpha)$ we define
\be \label{2.8i}
D^H_j=
[Q_v^{(1/2)}]^3\sum_{b(e)=v} E_j(e) \pi_J(v)[(\underline{A}(e))_{JK} 
\phi_K(f(e))-
\phi_J(v)]
\ee
\item[C.4] {\it Hamiltonian Constraint}\\
For any $v\in V(\alpha)$ we define
\ba \label{2.8j}
H^H(v)&=&\frac{1}{2} [Q_v^{(1/2)}]^3 \pi_J(v) \pi_J(v)
+ \frac{1}{2} V_v^{1/2} U(\phi(v))
+\frac{1}{2} [Q_v^{(1/2)}]^3[\sum_{b(e)=b(e')=v} 
\times\nonumber\\ && \times E_j(e) E_j(e')
[(\underline{A}(e))_{JK}\phi_K(f(e))-\phi_J(v)]
[(\underline{A}(e'))_{JL}\phi_L(f(e'))-\phi_J(v)] 
\ea
where $U$ is a positive, gauge invariant function of the $\phi_I(v)$, 
called the potential term.
\end{itemize}

\subsubsection{The (Extended) Master Constraint}
\label{s2.3.5}

We now simply add all the various geometry and matter 
contributions\footnote{We suppress appropriate numerical coefficients 
which turn all the terms to be added dimensionless and such that 
in the semiclassical limit \cite{II} these terms have the same 
coefficients as in the classical constraints. More precisely, 
for each commutator between a holonomy and a power $r$ of the volume 
operator we should divide by $r\hbar Q_{{\rm GR}}^2$ and each 
contribution to either constraint comes with an additional factor 
of $1/Q_{{\rm sector}}^2$.}  
\ba \label{2.8k}
G_j(v) &:=& G^{GR}_j(v)+G^F_j(v)
\nonumber\\
\underline{G}_J(v) 
&:=& 
\underline{G}^{YM}_J(v) 
+\underline{G}^F_J(v) 
+\underline{G}^H_J(v) 
\nonumber\\
D_j(v) 
&:=& D_j^{GR}(v) + D_j^{YM}(v) + D_j^F(v) + D_j^H(v)
\nonumber\\
H(v) &:=& H^{GR}(v) + H^{YM}(v) + H^F(v)+ H^H(v) + \Lambda \sqrt{V_v}
\ea
where we have added a possible cosmological term in the last line 
and can now simply define the Master Constraint as  
\be \label{2.9}
\MC:=\sum_{v\in V(\alpha)} [G_j(v)^\dagger  
G_j(v))+\underline{G}_J(v)^\dagger \underline{G}_J(v)+ D_j(v)^\dagger 
D_j(v)+ H(v)^\dagger H(v)]
\ee
The master constraint is manifestly positive and we take 
as its self -- adjoint extension the Friedrich's extension.\\
\\
Several remarks are in order:
\begin{itemize}
\item[I.] {\it Difference with background dependent theories}\\
What is remarkable about all these formulas is that they are rather 
similar to the expressions familiar from (Hamiltonian) lattice gauge 
theory. For instance, on a regular cubic spatial lattice embedded in 
$\Rl^3$ with edge length $\epsilon$ with respect to the standard
Euclidean metric the classical continuum Yang -- Mills Hamiltonian
\be \label{2.10a}
H_{YM}=\frac{1}{2Q^2}\int_{\Rl^3}\; d^3x\; \delta_{ab} {\rm Tr}[E^a E^b
+B^a B^b]
\ee
where $B^a=\epsilon^{abc} F_{bc}/2$ is the magnetic field and $F$ the 
curvature of the Yang -- Mills connection,
would be discretised in terms of our lattice variables as
\be \label{2.10b}
H_{YM}=\frac{1}{2Q^2\epsilon}\sum_v \delta_{ab} {\rm Tr}[E(e^a_v) E(e^b_v)
+A(\beta^a_v) A(\beta^b_v)]
\ee
where the sum is over all vertices of the lattice, $e^a_v$ is the edge
in the $a-th$ direction beginning at $v$ and $\beta^a_v$ is the the 
plaquette loop in the $x^a=$const. plane beginning at $v$. 

This expression should be contrasted with the classical expression 
for the Master Constraint on a differential manifold $\sigma$
(we just consider the contribution of the Euclidean Hamiltonian constraint 
for illustrative purposes)
\be \label{2.10c}
\MC=\int_\sigma \;d^3x\; \frac{[{\rm Tr}(F_{ab} E^a 
E^b)]^2}{\sqrt{|\det(E)|}}
\ee
which on a cubic algebraic graph could look like\\
\\
\be \label{2.10d}
\MC=\sum_v [\sum_a {\rm Tr}(A(\beta^a_v) A(e^a_v) 
[A(e^a_v)^{-1},\sqrt{V_v}])]^2
\ee
Expression (\ref{2.10d}), in contrast to (\ref{2.10b}) does not contain 
information about a background metric (there is none), a UV regulator 
$\epsilon$ or even the topology of $\sigma$. As long as the algebraic 
graph is infinite, it can be embedded arbitrarily densely into any 
manifold $\sigma$ and therefore {\it no continuum limit has to be taken}.
The theory is therefore UV finite.
\item[II.]
As we will see,
the kernel of the Master Constraint defines the states which are invariant 
under internal gauge transformations and, when embedded, under spacetime 
diffeomorphisms\footnote{The symmetries generated by the Hamiltonian and 
spatial diffeomorphism constraint have the interpretation of spacetime 
diffeomorphisms only when the equations of motion hold.} of GR. 
This is due to the simple fact that $\MC$ vanishes if and only if the
individual constraints hold\footnote{The proof of this statement is 
trivial for the case that zero is only in the point spectrum of some set
of (not necessarily self-adjoint) constraints $C_I$. Namely, $C_I\psi=0$ 
for all $I$ obviously implies $\MC\psi=0$ where $\MC=\sum_I C_I^\dagger 
C_I$. Conversely, $\MC\psi=0$ implies $<\psi,\MC\psi>=\sum_I || C_I 
\psi||^2=0$, hence $C_I \psi=0$ for all $I$. The general case is treated 
in complete detail in the first reference of \cite{Test}.} 
\item[III.]
In order to see that the solutions of the Master constraint are, in 
particular, what one 
intuitively expects of spatially diffeomorphism states that one can 
construct in the 
embedding dependent context \cite{ALMMT} one must embed those solutions.
At this point, the exact solutions of the Master Constraint in the new AQG 
context have not yet been constructed. However, one can perform tests that 
support our expectations. First of all, using coherent states one can show 
that the semiclassical limit of $\MC$ is correct. Next, approximate 
solutions to the Master constraint are coherent states which are peaked on 
the constraint hypersurface of the classical phase space and one can 
verify that the action of the diffeomorphism group derived in \cite{ALMMT}
leaves the state approximately invariant. Finally, one can try to improve 
the discretisations used in the above formulae which only uses next 
neighbour terms, to all neighbour terms in order to obtain a non -- 
anomalous quantum algebra on the abstract graph. This could be done, 
for instance by the method of perfect actions \cite{Hasenfratz}.  
\item[IV.]
As already mentioned, it is tempting to drop the spatial 
diffeomorphism constraint from our analysis because at the abstract graph 
level no diffeomorphisms can be defined. However, that is inconsistent as 
it does not correctly reduce the degrees of freedom as required by the 
spatial diffeomorphism constraint, because the abstract theory and the 
embedded theory should be in one to one correspondence as far as the 
physical degrees of freedom are concerned and when 
embedding the abstract graph, the diffeomorphism group acts non -- 
trivially. See \cite{Corichi} for a more detailed discussion.
\end{itemize}

\section{Semiclassical Analysis}
\label{s3}

We review elements of \cite{GCS,STW,Complexifier} which can be consulted 
for more details.\\
\\
We want to show that AQG is a canonical quantisation of classical General 
Relativity including matter. The canonical formulation of classical GR 
in the form we need it is reviewed for instance in \cite{books}. To begin 
with, the classical theory is formulated on manifolds diffeomorphic to
$\Rl\times \sigma$ where $\sigma$ is a three manifold of arbitrary 
topology. Thus, we must choose a differential manifold $\sigma$ and 
embed the fundamental algebraic graph $\alpha$ into $\sigma$. Its image 
will be called $\gamma:=X(\alpha)$. Notice that any three manifold admits
an infinite number of triangulations by tetrahedra or cubes and 
the graphs dual to such triangulations are simplicial (all vertices are 
four valent) or cubic (all vertices are six valent) respectively. Thus we 
focus on simplicial or cubic algebraic graphs.
If there are topological 
obstructions to embed the total $\alpha$ into $\sigma$ then we delete 
suitable parts of it until it can be embedded. We will then simply not 
excite the corresponding edges in the coherent state in what follows 
so that those edges drop out of all formulas (the coherent states are 
replaced by the function equal to one). An example is when $\sigma$ 
is compact so that embedding the infinite graph would lead to accumulation 
points.

\subsection{Gravity and Yang -- Mills Sector}
\label{s3.1}

We will choose embeddings $X$ such that $\gamma$ is dual to a certain 
triangulation $\gamma^\ast$. Thus, for each 
$X(e)$ there is a face $S_e$
in $\gamma^\ast$ which intersects $\gamma$ only in an interior point 
$p_e$ of both $S_e$ and $X(e)$. For each $x\in S_e$ we choose a path 
$\rho_e(x)$ which starts in $b(X(e))$ along $X(e)$ until $p_e$ and then 
runs within $S_e$ until $x$. Next, choose a classical $G-$connection 
$A_0$ and a Lie$(G)$ valued vector density $E_0$ of weight one. With 
the help of these data we define the quantities
\ba \label{3.1}
A_0(e)&:=& A_0(X(e)):={\cal P}\exp(\int_{X(e)} A_0)
\\
E_0(e)&:=& \int_{S_e} \epsilon_{abc} dx^a\wedge dx^b A_0(\rho_e(x))
(E_0)^c(x) A_0(\rho_e(x))^{-1}
\ea
which we will refer to as holonomies and electric fluxes respectively.

As one can show \cite{GCS}, if the classical theory is equipped with the 
following Poisson brackets
\be \label{3.2}
\{(A_0)_a^j(x),(A_0)_b^k(y)\}=\{(E_0)^a_j(x),(E_0)^b_k(y)\}=0,\;\;
\{(E_0)^a_j(x),(A_0)_b^k(y)\}=Q^2 \delta^a_b \delta_j^k \delta(x,y)
\ee
where $Q^2$ is the coupling constant ($G=Q^2/(8\pi)$ is Newton's constant
in GR), then the quantities (\ref{3.1}) satisfy
\ba \label{3.3}
\{A_0(e),A_0(e')\} &=& 0
\\
\{(E_0)_j(e),A_0(e')\} &=& \delta_{e,e'} \tau_j/2 A_0(e)   
\nonumber\\
\{(E_0)_j(e),(E_0)_k(e')\} &=& -Q^2 \delta_{e,e'}f_{jkl} (E_0)_l(e)
\nonumber
\ea
which precisely matches (\ref{2.1}). Hence, our kinematical algebra 
$\mathfrak{A}$ can be regarded as the quantisation of the reduction of the 
classical Poisson algebra to the quantities (\ref{3.1}) and we have 
considered a specific representation of $\mathfrak{A}$. 

We now consider coherent states. To that end we construct elements
of the complexification $G^\Cl$ of $G$ by
\be \label{3.4}
g_{e;(A_0,E_0)}:=\exp(i E_0(e)/a_e^2) A_0(e)
\ee
where we have introduced a parameter $a_e$ which may depend 
on $e$ whose dimension is such that $E_0(e)/a_e^2$ is dimensionfree
We now consider 
for $t>0$ and $g\in G^\Cl$
\be \label{3.5}
\Psi^t_g(h):=\sum_\pi \dim(\pi) e^{-t\lambda_\pi} \chi_\pi(gh^{-1})
\ee
Here the sum extends over all equivalence classes of irreducible 
representations of $G$ and  $\dim(\pi),\lambda_\pi,\chi_\pi$ respectively 
denote the dimension of $\pi$, eigenvalue of the Laplacian on $G$ when 
restricted to the representation space of $\pi$ and the character of 
$\pi$. For $G=SU(2)$ we have for instance
\be \label{3.5a}
\Psi^t_g(h):=\sum_j (2j+1) e^{-t j(j+1)/2} \chi_j(gh^{-1})
\ee
where the sum extends over all non negative half integers. 
The functions $\Psi^t_g$ are elements of $L_2(G,d\mu_H)$ and there is a 
measure $\nu^t$ on $G^\Cl$ such that the completeness relation holds
\be \label{3.6}
\int_{G^\Cl} d\nu(g) 
\frac{\overline{\Psi^t_g(h)} 
\Psi^t_g(h')}{||\Psi_g||\;||\Psi_{g'}||}=\delta_h(h')
\ee
where $\delta_h(h')=\Psi^0_h(h')$ is the $\delta-$distribution on $G$ 
with respect to the Haar measure. 

We now set 
$t_e:=\ell_P^2/a_e^2$ for gravity and 
\be \label{3.7}
\Psi_{e;(A_0,E_0)}(A):=\Psi^{t_e}_{g_{e;(A_0,E_0)}}(A(e))
\ee
and 
\be \label{3.8}
\Psi_{(A_0,E_0)}(A):=\otimes_{e\in E(\alpha)} \Psi_{e;(A_0,E_0)}(A)
\ee
It is important to keep in mind that (\ref{3.7}) is a state in the 
abstract graph Hilbert space, we just use all the data 
$\sigma,X,\gamma^\ast,\rho_e,A_0,E_0,a_e$ in order to construct specific 
elements of the abstract ITP Hilbert space. These states are coherent for 
our kinematical abstract algebra $\mathfrak{A}$ in the following sense:
Consider the ``annihilation operators''
\be \label{3.9}
g_e:=\exp(iE(e)/a_e^2) A(e)
\ee
Then our states satisfy\footnote{Up to a multiplicative factor which 
depends only on $t_e$ and tends to unity as $t_e\to 0$.}
\be \label{3.10}
g_e \Psi_{(A_0,E_0)}=g_{e;(A_0,E_0)} \Psi_{(A_0,E_0)}
\ee
that is, they are eigenstates of the annihilation operators. This is one 
of the defining properties of coherent states. These statements as well as  
other semiclassical 
properties such as peakedness properties are proved in \cite{GCS}. Of most 
importance for our purposes is that 
\be \label{3.10a}
<\Psi_{(A_0,E_0)},A(e) \Psi_{(A_0,E_0)}>=A_0(e),\;\;
<\Psi_{(A_0,E_0)},E(e) \Psi_{(A_0,E_0)}>=E_0(e)
\ee
up to terms which vanish faster than any power of $t_e$ as $t_e\to 0$.
Also the fluctuations are small, see \cite{GCS} for complete proofs. 

This holds for the gravity sector for which $E^a_j$ is dimensionfree while
$A_a^j$ has dimension cm$^{-1}$. This is why $\hbar Q^2=\ell_P^2$ has 
dimension of area. For Yang -- Mills theory $\underline{E}^a_J$ has 
dimension cm$^{-2}$ and $\underline{A}_a^J$ has dimension cm$^{-1}$ so 
that the Feinstrukturkonstante $\hbar Q^2$ is dimensionfree. Thus, for 
Yang -- Mills theory everything remains the same, the only difference 
being that the $a_e$ are now dimensionfree.\\
\\
For the mathematically inclined reader we mention that these states 
follow from an application of the complexifier framework 
\cite{Complexifier} which provides a constructive algorithm towards
coherent states. We define the positive operator, the complexifier
\be \label{3.11}
C:=-\frac{1}{2Q^2}\sum_{e\in E(\alpha)} \frac{1}{a_e^2} {\rm Tr}(E(e)^2)
\ee
and the $\delta-$distribution on the ITP Hilbert space ${\cal H}^\otimes$
\be \label{3.12}
\delta_A(A'):=\otimes \Psi^0_{A(e)}(A'(e))
\ee
Then 
\be \label{3.13}
\Psi_{(A_0,E_0)}=[e^{-C/\hbar} \delta_A]_{A(e)\to g_{e,(A_0,E_0)}}
\ee
and 
\be \label{3.14}
g_e=e^{-C/\hbar} A(e) e^{C/\hbar}
\ee
That is, the coherent states are nothing else than heat kernel evolutions 
of the $\delta-$distribution, analytically extended to complex group 
elements\footnote{Also coherent states constructed for the harmonic 
oscillator or free field theories fit into that scheme.}.

\subsection{Fermionic Sector}
\label{s3.2}

There is no such thing as a classical fermion. Only bilinear 
(commuting rather than anticommuting) expressions of the Grassman fields 
(``current densities'') have a classical interpretation. Hence, we are 
interested 
in semiclassical states which approximate the self -- adjoint quantities 
\be \label{3.20}
J^+_{MN}(v)=[\bar{\theta}_M(v)\theta_N(v)+\bar{\theta}_N(v)\theta_M(v)]/2,\;\;
J^-_{MN}(v)=[\bar{\theta}_M(v)\theta_N(v)-\bar{\theta}_N(v)\theta_M(v)]/(2i)
\ee
We will equivalently work with the non -- self adjoint currents 
$J_{MN}(v)=\bar{\theta}_M(v)\theta_N(v)$.
These satisfy the current algebra
\be \label{3.21}
[J_{MN}(v),J_{PQ}(v')] = \delta_{v,v'}[\delta_{NP} J_{MQ}(v)-\delta_{QM}
J_{PN}(v)]
\ee
We will construct semiclassical states for these currents, see \cite{CohSt}
for other proposals made in the literature. It will be sufficient to do this for
each $v$ separately. For each $v$ the Hilbert space is complex $2^N$  
dimensional 
while the number of currents is real $N^2$ dimensional where  
$N=2\dim(G)$ due to
the adjointness relation $J_{MN}^\ast=J_{NM}$. Since there are only 
$N$ fermionic degrees of freedom\footnote{Notice that $\bar{\theta}$ plays the
role of the conjugate momentum of $\theta$, hence one fermionic degree of
freedom counts for one configuration and one momentum degree of 
freedom.} 
$\theta_M$ which count $N$ complex degrees of freedom, we will not look for 
states which approximate 
all the currents but only the $N$ currents $J_{MM}=\bar{\theta}_M \theta_M$
and the remaining freedom in the states will be used in order to approximate
the phase of $\theta_M$ itself.

We notice that the Hilbert space at fixed $v$ is the span of states of the form
\be \label{3.22}
\Psi_{a,b}:=(a_1 + b_1 \theta_1) .. (a_N + b_N \theta_N)
\ee
for $a_k,b_k\in \Cl$ which are $2N$ complex degrees of freedom. In order to
reduce those to $N$ complex degrees of freedom we use the normalisation
$|a_k|^2/s+|b_k|^2=1$ for all $k$ which leaves us with $3N$ real degrees of
freedom. Here we have abbreviated $s=\hbar Q^2$. We compute
\be \label{3.23}
<\Psi_{a,b}, J_{MM} \Psi_{a,b}>=|a_M|^2,\;\;
<\Psi_{a,b}, \theta_M \Psi_{a,b}>=(-1)^{M-1} \bar{b}_M a_M
\ee
If we fix the expectation value of $J_{MM}$ to $j_M$ then 
$|a_M|^2=s(1-|b_M|^2)=j_M$ which shows that $0\le j_M\le s$ revealing that 
$\theta_M$ is a bounded operator\footnote{This follows already from the 
anticommutation relations: Since both $\theta\bar{\theta},\;\bar{\theta} \theta$
are positive operators while $\theta\bar{\theta}+\bar{\theta}\theta=s$ it
follows that $||\theta||,\;||\bar{\theta}||\le s$.}. Setting the expectation 
value of $\theta_M$ 
to be $z_M$ we see that $|z_M|=\sqrt{j_M[1-j_M/s]}$ is already fixed while
the phase is free and we have arg$(a_M)=(M-1)\pi+$arg$(b_M)+$arg$(z_M)$.
The fluctuation of $J_{MM}$ follows from the operator identities 
$\theta^2=\bar{\theta}^2=0$ so that 
$(\bar{\theta}\theta)^2=s\bar{\theta}\theta$
hence $<J_{MM}^2>-<J_{MM}>^2=j_M(s-j_M)$. The states (\ref{3.22}) obey the 
resolution of identity
\be \label{3.24}
{\bf 1}=\prod_{J=1}^N \int_0^1  r_J^{s-1} dr_J \int_0^{2\pi} 
\frac{d\phi_J}{2\pi} 
\int_0^{2\pi} \frac{d\varphi_J}{2\pi} |\Psi_{a,b}><\Psi_{a,b}|
\ee
where $r_J=|a_J|^2/s,\;\phi_J=$arg$(a_J)-$arg$(b_J),\;
\varphi_J=$arg$(a_J)+$arg$(b_J)$.     

In contrast to the semiclassical states defined for the gauge and gravitational
sector, the states $\Psi_{a,b}$ defined for one vertex have large fluctuations.
This is due to the fact that what we should consider are not current densities
but rather currents, that is, expressions of the form 
$J_{MM}(B)=\sum_{v\in B} J_{MM}(v)$ where $B\subset V(\alpha)$. Then
the relative fluctuation with respect to the states 
\be \label{3.25}
\Psi=\otimes_{v\in V(\alpha)} \Psi_{a_v,b_v}
\ee
is given by 
\be \label {3.26}
\frac{<\Psi, J_{MM}(B)^2 \Psi> - <\Psi, J_{MM}(B)\Psi>^2}
{<\Psi, J_{MM}(B)\Psi>^2} 
=\frac{\sum_{v\in B} j_M^v(s-j_M^v)}{[\sum_{v\in B} j_M^v]^2}\propto 
\frac{1}{|B|}
\ee
if $j_v\approx j$ is not varying too much over $B$. We see that macroscopic 
currents have very small fluctuations. 

Geometrically, the relation between the components of a Weyl spinor $\xi_M(x)$ 
(which transforms as a scalar under spatial diffeomorphisms)
as it appears in the classical action and the $\theta_M(v)$ 
is given by the formula \cite{QSD}
\be \label{3.27}
\root 4\of{\det(q)(x)} \xi_M(x):=\sum_{v\in V(\alpha)} \theta_M(v)
\sqrt{\delta(X(v),x)} 
\ee
where the three metric $q_{ab}$ has appeared explicitly and $X$ is the embedding
again. The square root of the $\delta-$ distribution matches the density weight
of the equation. Notice that $\xi$ vanishes away from the vertices of the 
embedded graph. Using $\sqrt{\delta(x,y)\delta(x,z)}:=\delta_{x,y}\delta(x,z)$
it is easy to see that we have for the spatially diffeomorphism invariant 
quantity
\be \label{3.28}
\int_\sigma d^3x \sqrt{\det(q)}(x) \bar{\xi}_M(x)\xi_M(x)=\sum_{v\in V(\alpha)}
\bar{\theta}_M(v)\theta_M(v)
\ee

\subsection{Higgs Sector}
\label{s3.3}

For the Higgs Sector we can construct coherent states of a more 
traditional type. Given a classical canonical pair $(\phi_0)_I,(\pi_0)_I$
equipped with the Poisson brackets
\be \label{3.50}
\{(\phi_0)_I(x),(\phi_0)_J(y)\}=\{(\pi_0)_I(x),(\pi_0)_J(y)\}=0,\;
\{(\pi_0)_I(x),(\phi_0)_J(y)\}=Q^2\delta_{IJ} \delta(x,y)
\ee
we consider for each vertex $X(v)$ of the embedded graph the variables 
$(\phi_0)_I(v):=(\phi_0)_I(X(v))$ and 
$(\pi_0)_I(v):=\int_{C_v} d^3x (\pi_0)_I(x)$ where $C_v$ is the cell of 
the dual cell complex $\gamma^\ast$ which contains $v$. These variables 
induce the Poisson brackets
\be \label{3.51}
\{(\phi_0)_I(v),(\phi_0)_J(v')\}=\{(\pi_0)_I(v),(\pi_0)_J(v')\}=0,\;
\{(\pi_0)_I(v),(\phi_0)_J(v')\}=Q^2 \delta_{IJ} \delta_{v,v'}
\ee
which is compatible with (\ref{2.3c}). If we take the Higgs field to be 
dimensionfree then $\hbar Q^2$ has dimension cm$^2$ and the $\phi_0(v)$ 
have dimension cm$^2$. Hence we introduce parameters $L_v$ of dimension 
cm and from those annihilation operators 
\be \label{3.52}
a_I(v):=\frac{1}{\sqrt{2}}[\phi_I(v)-i \pi_I(v)/L_v^2] 
\ee
and complex numbers
\be \label{3.53}
z_I(v):=\frac{1}{\sqrt{2}}[(\phi_0)I(v)-i (\pi_0)I(v)/L_v^2] 
\ee
From these we construct  
\be \label{3.54}
\Psi^t_z=e^{-|z|^2/2} e^{z a} \Psi_0,\;\;\Psi_0(x)=e^{-x^2/t_v}/\sqrt{2\pi 
t_v}
\ee
and then
\be \label{3.55}
\Psi_{(\phi_0,\pi_0)}:=\otimes_{v\in V(\alpha),I} \Psi^{t_v}_{z_I(v)}
\ee
where $t_v=\hbar Q^2/L_v^2$.\\ 
\\
Remark:\\
In contrast to the LQG representation which is necessarily discontinuous 
in the edge labels of the holonomy operators so that the connection operator 
(smeared over one dimensional paths) does not exist, in AQG we may indeed define
such a representation. We simply define a new algebra by 
\be \label{3.56}
[A_j(e),A_k(e')]=[E_j(e),E_k(e')]=0,\;\;[E_j(e),A_k(e')]=i\hbar Q^2
\delta_{jk} \delta_{ee'}
\ee
where now both $E(e),A(e)$ are Lie$(G)$ valued. This $^\ast-$algebra is
represented on the infinite tensor product of Hilbert spaces, one for each
edge, of the Hilbert space $L_2(\Rl^{\dim(G)},d^{\dim(G)}x)$ on which $A_j(e)$ 
and $E_j(e)$ respectively act by multiplication and derivation respectively by
$x_j$. Such a representation is forbidden in LQG because one needs to relate the
Hilbert spaces defined for different (infinite) graphs to each other in such a
way as to respect the relations $A(e_1\circ e_2)=A(e_1)+A(e_2),\;A(e^{-1})=-
A(e)$. One can easily see that there is no underlying cylindrically consistent
measure underlying such a Hilbert space because the divergence of the electric
flux operator with respect to such a measure is not $L_2$, see e.g.
\cite{Sahlmann}. ITP Hilbert spaces have no underlying measure, however,
now the definition of the inner product between vectors belonging to two
different ITP's based on different graphs becomes problematic, see \cite{GCS}.
In AQG there is only one graph and therefore the problem disappears.

One would then define $A_0(e)=\int_e A_0$ and then define harmonic oscillator
type of coherent states just as in (\ref{3.52}) -- (\ref{3.54}). At least one
could do that for the matter gauge fields such as the Maxwell field for 
which
oscillator type of coherent states were actually invented. For gravity one 
might
want to stick with the algebra of section \ref{s2} in order to keep the
discreteness of the spectrum of geometrical operators.

\section{(One) Semiclassical Limit of the Master Constraint}
\label{s4}

In what follows we summarise the result of \cite{II} where a semiclassical calculation for the extended algebraic master constraint operator based on a cubic 
algebraic graph is presented.
The calculation makes use of the following approximation: We substitute 
the gauge group $SU(2)$ by $U(1)^3$. This is of course incorrect, however, 
the results of \cite{III,GCS} together show that the results of the 
exact 
non Abelean calculation match precisely the results of the Abelean 
approximate calculation, provided one substitutes in the result of the 
approximate expectation value calculation every Abelean holonomy and 
electric flux by the corresponding non -- Abelean quantity. More 
precisely, the symplectic structure (\ref{3.2}) does not know whether
we are given a $SU(2)$ or $U(1)^3$ gauge theory, the phase space is the 
same, only if we add the constraints do we get this additional 
information. Hence, we may use a point $(A_0,E_0)$ in the common phase 
space of both theories. In order to carry out the approximate calculation,
the non -- Abelean operators 
$({\rm Tr}(\tau_j A(e)), {\rm Tr}(\tau_j E(e)))$ 
are replaced by the Abelean ones 
$(h^j(e), p_j(e))$ where $h^j(e)$ corresponds to the holonomy of
the j-th copy of $U(1)$ and likewise for the electric field. Note, that on purpose we introduce new letters for the holonomy and the electric flux in order to distinguish more easilier whether we are talking about $U(1)^3$ or $SU(2)$.
Then, after 
the expectation value is calculated 
one replaces the classical $U(1)^3$ terms 
$((h_0)^j(e), (p_0)_j(e))$ by   
$({\rm Tr}(\tau_j A_0(e)), {\rm Tr}(\tau_j E_0(e)))$.
The result of that calculation turns out to be exactly the same as if 
directly doing the non Abelean calculation, of course only to zeroth order 
in $\hbar$. The advantage of this indirect calculation is that it is much 
easier to perform.

In order to do this, all we have to do is to change the coherent states 
from those for $SU(2)$ to those of $U(1)^3$. This is rather easy:
Consider the state
\be \label{4.1}
\Psi^t_g(h):=\sum_{n\in \Zl} e^{-tn^2/2} (gh^{-1})^n
\ee
where $g\in \Cl-\{0\}=U(1)^{\Cl}$ and $h\in U(1)$. The function 
(\ref{4.1}) is an element of $L_2(U(1),d\mu_H)$. We set for $j=1,2,3$
\be \label{4.2}
g^j_{e;(A_0,E_{0})}:=e^{E^j(e)/a_e^2} e^{i\int_e A_0^j}
\ee
and with $t_e:=\ell_p^2/a_e^2$
\be \label{4.3}
\Psi^{\{t_e\}}_{\alpha,(A_0,E_0)}:=\otimes_{e\in E(\alpha)} \otimes_{j=1}^3
\Psi^{t_e}_{\alpha,g^j_{e;(A_0,B_0)}}
\ee
For simplicity and since this will not affect the final result, we choose the same $t_e=:t$ for each edge. Moreoever we will introduce the shorthand $m:=(A_0,E_0)$ for the phase space point. The coherent states are then denoted by 
\be
\Psigeng=\otimes_{e^\in E(\alpha)} \otimes_{j=1}^3
\Psi^{t}_{\alpha,g_{e},m}
\ee
Requiring the graph $\alpha$ to have cubic symmetry we know that each vertex is six-valent. We label these six edges by $e^{\sigma}_J$, whereby $\sigma\in\{+,-\}$ depending on the orientation with respect to the vertex $v$ and $J\in\{1,2,3\}$. For more details see \cite{II}. Let us introduce the following notation for the $U(1)$-holonomies and electric fluxes
\ba
h_{J\sigma jv}:&=&h^j_{e^{\sigma}_J(v)}\nonumber\\
p_{J\sigma jv}:&=&p_j^{e^{\sigma}_J(v)}\nonumber\\
h_{\In\sn\Jn\sn^{\prime}j_0v}&:=&h_{\In\sn j_0v}h_{\Jn\sn^{\prime}j_0v}h^{-1}_{\In\sn j_0v}h^{-1}_{\Jn\sn^{\prime}j_0v}
\ea
For the considered algebraic graph of cubic symmetry the algebraic master constraint operator denoted by $\MCO$ has the following form
\ba
\MCO&=&\sum\limits_{v\in V(\gamma)}\MCO_v\nonumber\\
\MCO_{v}&=&\sum\limits_{\elln=0}^{3}\op{C}_{\elln,v}^{\dagger}\op{C}_{\elln,v}\nonumber\\
\op{C}_{0,v}&=&\sum\limits_{\In\Jn\Kn}\sum\limits_{\sn=+,-}\sum\limits_{\sn^{\prime}=+,-}\sum\limits_{\sn^{\prime\prime}=+,-}\frac{4}{\kappa}\epsilon^{\In\Jn\Kn}
\holloop{\In \sn^{\prime}}{\Jn}{\sn^{\prime\prime}}{\elln}{v}\hol{\Kn}{\sn}{\elln}{v}\frac{1}{i\hbar}\left[\holin{\Kn}{\sn}{\elln}{v},\op{V}^{\frac{1}{2}}_{\gamma,v}\right]\nonumber\\
\op{C}_{\elln,v}&=&\sum\limits_{\In\Jn\Kn}\sum\limits_{\sn=+,-}\sum\limits_{\sn^{\prime}=+,-}\sum\limits_{\sn^{\prime\prime}=+,-}\frac{4}{\kappa}\epsilon^{\In\Jn\Kn}\epsilon_{\elln \mn\nn}
\holloop{\In \sn^{\prime}}{\Jn}{\sn^{\prime\prime}}{\mn}{v}\hol{\Kn}{\sn}{\nn}{v}\frac{1}{i\hbar}\left[\holin{\Kn}{\nn}{\sn}{v},\op{V}^{\frac{1}{2}}_{\gamma,v}\right]
\ea
where the volume operator of the cubic graph expressed in terms of right invariant vector fields $\op{X}^{e^{\sigma}_J}_j:=\Ygen=i\hgen\partial/\partial\hgen$is given by
\be
\op{V}_{\alpha,v}=\lp^3\sqrt{\left|\epsilon^{jkl}\left[\frac{\Y{1}{+}{j}{v}-\Y{1}{-}{j}{v}}{2}\right]\left[\frac{\Y{2}{+}{k}{v}-\Y{k}{-}{j}{v}}{2}\right]\left[\frac{\Y{3}{+}{l}{v}-\Y{3}{-}{l}{v}}{2}\right]\right|}
\ee
with its corresponding eigenvalue 
\be
\lambda^{\frac{1}{2}}(\{\ngen\})=\lp^3\left(\sqrt{\left|\epsilon^{jkl}\left[\frac{\n{1}{+}{j}{v}-\n{1}{-}{j}{v}}{2}\right]\left[\frac{\n{2}{+}{k}{v}-\n{k}{-}{j}{v}}{2}\right]\left[\frac{\n{3}{+}{l}{v}-\n{3}{-}{l}{v}}{2}\right]\right|}\right)^{\frac{1}{2}}
\ee
Our task is now to show that the expectation value
\be \label{4.4}
\frac{<\Psigeng,\MCO\Psigeng>}{||\Psigeng||^2}
\ee
coincides with 
the classical $U(1)^3$ Master Constraint 
\be \label{4.5}
\MC[m]=\Big\{\int_\sigma \; d^3x\; 
\frac{\delta^{jk} C_j C_k+q^{ab} C_a C_b+ 
C^2}{(\sqrt{\det(q)})^3}(x)\Big\}[m]
\ee
evaluated at the point $m=(A_0,E_0)$ in the classical phase space in the limit $\hbar\to 0$. 
 Here the following functions were defined
(we drop the subscript ``0'')
\be \label{4.6}
C_j = \partial_a E^a_j 
\quad
C_a = F_{ab}^j E^b_j
\quad
C = \epsilon^{abc}[F_{ab}^j+\epsilon_{jkl} K_a^j K_b^k] e_c^l 
\ee
where 
\ba \label{4.7}
F_{ab}^j &=& 2\partial_{[a} A_{b]}
\quad
E^a_j = |\det((e_b^k))| e^a_j,\;\;e^a_j e_b^j=\delta^a_b,\;\; e^a_j 
e_a^k=\delta_j^k
\nonumber\\
q_{ab} &=& e^j_a e^j_b
\quad
K_a^j = A_a^j-\Gamma_a^j
\nonumber
\ea
and where $\Gamma_a^j$ is the spin connection of the co -- triad $e_a^j$,
that is, $D_a e_b^j=\partial_a e_b^j -\Gamma_{ab}^c e_b^j+\epsilon_{jkl} 
\Gamma_a^k e_b^l=0$, $\Gamma_{ab}^c$ are the Christoffel symbols 
determined by the three metric $q_{ab}$. 

Thes are the $U(1)^3$ quantities, the $SU(2)$ quantities are defined in 
exactly the same way, only the two following functions need to be changed
to
\ba \label{4.8}
C_j &\to & \partial_a E^a_j +\epsilon_{jkl} A_a^k E^a_l 
\\
F_{ab}^j & \to & \partial_{[a} A_{b]}^j+\epsilon_{jkl} A_a^k A_b^l
\nonumber
\ea 
That the $U(1)^3$ calculation has anything to do with the result of the 
exact $SU(2)$ calculation relies on the fact, established in \cite{III} 
that the $SU(2)$ volume operator can be semiclassically expanded in 
terms of polynomials of flux operators plus $\hbar$ corrections. 
However, as shown in \cite{GCS}, to zeroth order in $\hbar$, expectation 
values of polynomials of holononomy and flux operators agree in  
$U(1)^3$ and $SU(2)$ calculations and also extends to operators of type 
$Q^{(r)}$ as shown in \cite{ST}.
As long as we arrive at the correct classical $U(1)^3$ master constraint in the leading order of the expectation value calculation we are also qualitatively done for $SU(2)$.\\
In \cite{II} we proof that the expectation value of the algebraic master constraint operator associated to a graph of cubic topology yields in the leading order
\ba
\lefteqn{\frac{\langle\Psigen\,|\,\MCO\,|\,\Psigen\rangle}{||\Psigen||^2}}\nonumber\\
&=&
\sum\limits_{v\in V(\alpha)}\lefteqn{\frac{\langle\Psigen\,|\,\MCO_{v}\,|\,\Psigen\rangle}{||\Psigen||^2}}\nonumber\\
&=&
\sum\limits_{v\in V(\alpha)}\sum\limits_{\In\Jn\Kn}\sum\limits_{\Int\Jnt\Knt}\sum\limits_{\sn=+,-}\sum\limits_{\snt=+,-}\epsilon^{\In\Jn\Kn}\epsilon^{\Int\Jnt\Knt}\lb(\delta_{\mn,\nn}\delta_{\mnt,\nnt}
+\sum\limits_{\elln=1}^3 \epsilon_{\elln \mn\nn}\epsilon_{\elln \mnt\nnt}\rb)\nonumber\\
&&
\Big\{
\lb(\frac{4a^{\frac{3}{2}}|\det(\pmm)|^{\frac{1}{4}}}{\kappa\hbar}\rb)^2\lb(sT\rb)^2e^{+i\sum\limits_{\vt\in V}\sum\limits_{(J,\sigma,j)\in L}\phigen\Del}\nonumber\\
&&\lb(f^{(1)}_{\frac{1}{8}}(1)\rb)^2\lb(\sgn(\sn)\qmm_{\Kn\nn}\rb)\lb(\sgn(\snt)\qmm_{\Knt\nnt}\rb)\Big\}+O((sT/t)^2)
\ea
The leading has to be understood as follows. The coherent are labelled with the so called classicality parameter $t\propto \hbar$. Hence, the limit $\lim\limits_{t\to 0}$ corresponds to the limit $\hbar\to 0$ and is the limit in which the expectation value should agree with the classical quantities to approximate.
We show in detail in \cite{II} that the result of the expectation value of $\MC$ in the limit $t\to 0$ above can be identified with the classical discretised master constraint associated to a cubic lattice, denoted by $\MC^{cubic}$ from now on. Furthermore, we proof that $\MC^{cubic}$ agrees indeed with the classical continuum expression of the master constraint $\MC$ in eqn (\ref{4.5}) when shrinking the parameter intervall length $\epsilon$ to zero. This can be summarised in the following equation
\be
\fbox{$\vspace{0.2cm}\frac{\langle\Psi^t_{\alpha,m}\,|\,\MCO\,|\,\Psi^t_{\alpha,m}\rangle}{||\Psi^t_{\alpha,m}||^2}=\sum\limits_{v\in V(\alpha)}\frac{\langle\Psigen\,|\,\MCO_{v}\,|\,\Psigen\rangle}{||\Psigen||^2}{=\atop\lim\limits_{t\to 0}}\MC^{cubic}[m]{=\atop\lim\limits_{\epsilon\to 0}}\MC[m]\vspace{0.2cm}$}
\ee
Consequently, with the calculation done in \cite{II} we have shown that Algebraic Quantum Gravity is a theory of quantum gravity which  has the same infinitesimal generators as General Relativity. Thus, the problem whether the semiclassical sector includes General Relativity, that is still unsolved within the framework of Loop Quantum Gravity, is significantly improved  in the context of Algebraic Quantum Gravity.
Additionally, we discuss the next-to-leading order term of the expectation value which can be interpreted as fluctuations of $\MCO$. It turns out that these next-to-leading order contributions are finite. For a more detailed discussion see \cite{II}.
\\
Let us close this section with some remarks concerning the details of the analysis in \cite{II}:
\begin{itemize}
\item[1.]
In \cite{II} we only considered the gravitational sector, however, the 
techniques used there carry over to all standard matter coupling. 
\item[2.]
In \cite{II} we also dropped the piece corresponding to the 
quantum Gauss constraint because it is just a linear 
combination of flux operators for which the correct classical limit was 
established in \cite{GCS} already.\\
\item[3.]
In \cite{II} we only considered the Euclidean part of the 
Hamiltonian constraint. The Lorentzian piece cannot be correctly 
produced using $U(1)^3$ because the classical identity 
$\{H_E^{(1)},V\}=\int_\sigma d^3x K_a^j E^a_j$ for $SU(2)$ on which 
(\ref{2.8}) relies fails to hold. However, again the results of 
\cite{II,III} show that the correct $SU(2)$ calculation does reproduce 
the correct classical limit. \\
\end{itemize}
\section{Computational AQG and Quantum Gauge Fixing}
\label{s5}

The fact that the Master Constraint has the correct classical limit in AQG
is a strong indicator that the theory has the correct classical limit 
because the Master constraint determines both the physical Hilbert space 
and the quantum observables which are required to preserve the physical 
Hilbert space. 
Ideally, in order to establish this one needs to compute the physical Hilbert 
space, construct the gauge invariant quantum observables and define a 
dynamics among those\footnote{Notice that in background independent 
theories there is no natural Hamiltonian, the Hamiltonian constraint 
is constrained to vanish and observables need to commute with it.
Hence the Hamiltonian constraint is unsuitable to define dynamics.
Extra work is required in order to define evolution among observables, see 
below.}.

As far as the first task is concerned, this can be done as follows: 
As is well -- known (see the first reference of \cite{Test} for all 
details), given a self -- adjoint operator $\MC$ on a separable Hilbert 
space $\cal H$, there is a unitarily equivalent representation of 
$\MC$ on 
a direct integral Hilbert space 
\be \label{5.1}
{\cal H}\cong {\cal H}^\oplus=\int^\oplus_{{\rm spec}(\MC)} 
\;d\mu(\lambda) \; {\cal H}^\oplus_\lambda
\ee
where $\mu$ is a spectral measure for $\MC$, spec$(\MC)$ denotes the 
spectrum of $\MC$ and the separable Hilbert spaces ${\cal 
H}^\oplus_\lambda$ are the 
generalised eigenspaces of $\MC$ in the following sense: Given 
$\Psi\in {\cal H}$ we can represent it as a system of vectors
$(\Psi_\lambda)_{\lambda\in {\rm spec}(\MC)}$ where 
$\Psi_\lambda\in {\cal H}^\oplus_\lambda$. Then $\MC\Psi$ is represented 
as the system $(\lambda \Psi_\lambda)$. The inner product is given by
\be \label{5.2}
<\Psi,\Psi'>=\int\; d\mu(\lambda)\; <\Psi_\lambda,\Psi'_\lambda>_{{\cal 
H}_\lambda}
\ee
This is really nothing else than a generalisation of the Fourier transform 
to an arbitrary self -- adjoint operator, the dimension of ${\cal 
H}^\oplus_\lambda$ has the interpretation of the multiplicity of 
$\lambda$. 

The physical Hilbert space is the kernel of $\MC$, that is
${\cal H}_{phys}={\cal H}^\oplus_0$. The construction of $\mu$ and ${\cal 
H}^\oplus_\lambda$ requires the detailed knowledge of the spectrum of 
$\MC$ but otherwise there is a clean algorithm for how to obtain these 
structures which are unique up to unitary equivalence\footnote{There are 
some remaining ambiguities associated with the fact that equalities hold 
up to measure $\mu$ zero sets. For how to fix them, see \cite{Test}.}.
While the assumption of separability does not apply to the ITP Hilbert 
space ${\cal H}^\otimes$, there is no problem because $\MC$ preserves all 
the strong equivalence class Hilbert spaces. This follows from the 
fact that $\MC$ is a countable sum of operators each of which 
changes only a finite number of entries in a vector of the form 
$\otimes_f$, hence we get a countable sum of vectors in the same 
equivalence class, which remains normalisable if $\otimes_f$ is in 
the domain of $\otimes_f$. Hence, we can apply the 
direct integral decomposition to each of these separable Hilbert spaces 
separately.

The quantum observables are the self -- adjoint operators on ${\cal 
H}_{phys}$. This is mathematically sufficient but we are interested  
in those observables with a classical interpretation, that is, those which
can be defined on the kinematical Hilbert space ${\cal H}^\otimes$, which 
have a classical limit in the sense of our coherent states and 
which preserve the eigenspaces ${\cal H}^\oplus_\lambda$. As can be shown, 
a function $F$ on the classical phase space is an observable provided that
$\{F,\{F,\MC\}\}_{\MC=0}=0$, see \cite{M1}. A systematic way to construct 
such observables is via the partial observable Ansatz due to Rovelli
\cite{Rovelli}, see  
\cite{B1,B2,Reduced} and references therein for recent improvements 
concerning the technical implementation. This is a classical framework 
which, given a set of constraints $C_I$, a set of phase space functions 
$T_I$ subject to $\det((\{C_I,T_J\}))\not=0$, a set of real numbers 
$\tau_I$ in the range of the $T_I$, and a phase space 
function $f$, constructs an observable $F_{f,T}^\tau$ as a power series 
in the variables $\tau_I-T_I$. Hence
\be \label{5.3}
F_{f,T}^\tau=f+\sum_I (\tau_I-T_I) f_I+ \sum_{I,J}
(\tau_I-T_I)(\tau_J-T_J) f_{IJ} +....
\ee
for certain phase space functions $f_I, f_{IJ},..$ which can be explicitly 
constructed. Physically, the $T_I$ are gauge fixing 
functions and if we evaluate $F_{f,T}^\tau$ at a point in phase space for 
which $T_I=\tau_I$ then $F_{f,T}^\tau=f$. The meaning of the real 
parameters $\tau_I$ is that each of them defines a physical time evolution 
because $F_{f,T}^\tau$ is an observable for each value of the $\tau_I$. 
One can also show that the evolution in $\tau_I$ is generated by a 
physical Hamiltonian $H_I(\tau)$ which in general, however, will be $\tau$ 
dependent. Of course, the Hamiltonian should be bounded from below and 
should reduce to the Hamiltonian of the standard model when the metric 
is close to the Minkowski metric plus small fluctuations.

Unfortunately, all of that framework is purely classical and difficult 
to quantise because the expression for $F_{f,T}^\tau$ faces, in general, 
severe operator ordering problems. In order to sidestep these problems it 
would be desirable to have a more direct procedure at one's disposal 
in order to generate a physical Hamilonian. One way to do this is via
the Brown -- Kuchar mechanism based on a phantom field \cite{BK}. By 
choosing a suitable action for the phantom field, one can generate a 
physical Hamiltonian which reduces to the one of the remaining matter 
and gravity when the phantom field distribution is homogeneous. 
That Hamiltonian is explicitly $\tau-$independent and non -- 
negative. Classical physical 
observables can be constructed as well which suffer from less severe 
ordering problems. This is due to the fact that the phantom field allows 
for an explicit deparameterisation of the entire physical system. One 
might think that  
the drawback of this is that the phantom field is a scalar which 
has not been observed, but actually there is no problem because the 
phantom field is pure gauge anyway.

Thus we see that, apart from the technical problem to compute all of these 
quantities, there is a clear conceptual path for how to do physics with 
AQG. For instance, the physical Hamiltonian maybe used in order to select 
the true vacuum of the universe, a quantity that is ambiguous in the 
framework of quantum field theory on curved spacetimes. Furthermore, it 
will be used in order to compute physical scattering amplitudes. However,
in order to do so, we really need effective computational tools. The 
computation of the exact physical Hilbert space will be impossible due to 
the complexity of the theory so that we have to resort to approximations.
In a background independent and therefore necessarily non -- perturbative 
theory, only non -- perturbative tools are allowed. These are 
precisely the coherent states defined in section \ref{s3}: We will choose
a point in the classical phase space which 1. lies on the constraint 
surface of the classical Master constraint and 2. satisfies the gauge 
fixing conditions $T_I=\tau_I$ of our chosen functions $T_I$ (in our case 
essentially the phantom field). This means that these states are 
approximately physical states because the norm of the Master constraint
(equivalently its fluctuation) is close to zero and expectation values 
of physical observables $F_{f,T}^\tau$ effectively reduce to the 
expectation value of $f$, of course only to lowest order in the 
fluctuations of the $T_I-\tau_I$. 

One could call this approximation ``Quantum Gauge Fixing'' for the 
following reason: We are working at the level of the kinematical Hilbert 
space. We choose a state which is peaked on a point $m$ of the 
constraint surface and within its orbit $[m]$ on that point which 
corresponds to the gauge cut $T=\tau$. However, there are still 
fluctuations of all degrees of freedom involved, not only physical ones,
in particular in directions off the constraint surface and within the 
gauge orbit.
This is in contrast to gauge fixing before quantising. In a way we gauge 
fix after quantising by choosing appropriate states which suppress the 
fluctuations into the unphysical directions.
In the longer range, one has of course 
to answer the question how good this approximation is as compared to the 
exact calculation.

\section{Algebraic Quantum Gravity and Spin Foams}
\label{s6}

The algebraic or embedding independent setting proposed for AQG also
provides an interesting new perspective for the spin foam programme
\cite{spinfoam}. Spin foam models try to provide a path integral 
representation of LQG. Two of the most important tasks to be completed 
within the   
spin foam programme for 4D General Relativity\footnote{There are many 
promising results in 3D but this is hardly surprising since 2+1 
gravity is a TQFT. Most of the results in 3D rely on the TQFT structure 
and therefore do not carry over to the 4D case.} are 1. to make contact 
with the canonical theory
and 2. to remove the triangulation dependence of the models. 

In more detail, the spin foam models currently discussed in the 
literature start from a path integral that involves a constrained 
BF theory action. Classically, if one solves those so -- called 
simplicity constraints which impose that the $B$ field is the exterior
product of two Vierbeine, then one obtains the Palatini action and a 
topological term. In order to define the path integral mathematically 
one regularises it by choosing a triangulation and discretises the 
constrained BF theory on this triangulation. However, to the best 
knowledge of the authors, none of the spin foam models currently 
discussed has properly implemented the quantum simplicity constraints 
nor has dealt with the fact that the Palatini theory leads to second 
class constraints in the canonical formulation which has a non -- 
trivial effect on the path integral measure if the canonical and 
covariant theory are to compute the same thing. This is well known,
see for instance \cite{Henneaux} and has also been pointed out in 
\cite{Noui} for the spin foam context. 

The second issue has to do with the removal of the regulator, that is,
triangulation dependence. A natural idea would be to sum over 
triangulations, the choice of the weights being motivated by the group 
field theory formulation of spin foams \cite{spinfoam}. However, again 
to the present authors it is completely unclear how to make contact 
between the original path integral for the constrained BF theory which 
at least has a clear connection to the classical theory we want to 
quantise and the group field theory formulation. For instance, is it not 
more natural to study the infinite refinement limit of spin foam 
models and to look for critical points as in lattice gauge theory
or dynamical triangulations \cite{Ambjorn}?\\
\\     
We will now show that AQG offers a clean solution to both problems.
Indeed,
as advertised in \cite{M1}, the extended Master Constraint defines a new 
type of spin foam model which computes by means of the the rigging map 
heuristically\footnote{See the first reference of \cite{Test} for the 
rigorous definitions.} given by 
\be \label{6.1}
\eta:\;{\cal H}\to {\cal H}_{{\rm phys}};\;\psi\mapsto
\int_{\Rl} \; dt\; \exp(it \MC)\psi
\ee
the physical inner product 
\be \label{6.2}
<\eta(\psi),\eta(\psi')>_{{\rm phys}}:=
\int_{\Rl} \; dt\; <\psi,\exp(it \MC)\psi'>
=\int_0^\infty \; dt\; [<\psi,\exp(it \MC)\psi'>+
<\psi,\exp(-it \MC)\psi'>]
\ee
If the expression $<\psi,\exp(\pm it \MC)\psi'>$ is analytic in $t$ (for 
instance if $\psi$ or $\psi'$ are analytic vectors for $\MC$) then 
it can be considered as the analytic continuation $t\mapsto \mp i t$
in $t$ of the expression $<\psi,\exp(-t \MC)\psi'>$. Since $\MC$ is 
positive, the operators $\exp(-t\MC)$ are bounded for $t\ge 0$ 
(they form a contraction semi -- group) and have improved convergence 
properties as compared to the unitarities $\exp(\pm it \MC)$. 

Notice that $<\psi,\exp(-t \MC)\psi'>$ vanishes when $\psi,\psi'$ do not 
belong to the same sector of the ITP.  If we now 
write $\exp(-t\MC)=[\exp(-t\MC/N)]^N$ and insert $N-1$ resolutions
of unity $1_{{\rm sector}}=\sum_s |s><s|$ where $|s>$ denotes a 
countable orthonormal basis for the given sector then we arrive at a 
path integral formulation of the physical inner product. The 
orthogonality of the kinematical sectors carries over to their images 
under the rigging map. 

Let us restrict for the purpose of this article to the case that 
the semiclassical theories we want to quantise have compact $\sigma$.
The appropriate sector of the ITP is then based on the vector 
$\otimes_{{\rm\bf 1}}=\otimes_e\;{\rm\bf 1}$ where {\bf 1} is the 
constant function equal to one. 
An orthonormal basis for this sector is given by spin network 
functions defined over all finite subgraphs of the algebraic graph. Then
(\ref{6.2}) defines a concrete spin foam model of General 
Relativity {\it for which the issue of triangulation dependence is 
absent}. Notice that we may leave $N$ large but finite, the formula one 
obtains is exact for any $N$. Depending on the ``boundary states''
$\psi,\;\psi'$ and the value of $N$, the non vanishing 
contributions to the resulting sum will be over subgraphs of the 
algebraic graphs which reach a certain maximum size. This should be 
quite similar to the 3D model discussed in \cite{Perez}.
Details will follow in future publications.

\section{Conclusions and Outlook}
\label{s7}

Algebraic Quantum Gravity (AQG) offers a conceptually clear
and technically simpler approach to quantum gravity than Loop Quantum 
Gravity (LQG). The simplification occurs because in AQG one just has to 
deal with {\it one}, albeit countably infinite, algebraic graph while
in LQG one deals with an uncountably infinite number of finite and 
embedded graphs. 
In LQG this has the effect that the Hamiltonian constraint always refines 
the graph on which it acts while in AQG the algebraic graph is the finest 
possible one. The search for semiclassical states for such refining or 
graph changing operators has so far been unsuccessful. However, as we have 
indicated here and as will be shown in \cite{II,III}, the present 
semiclassical tools developed in \cite{GCS} are already sufficient to 
establish the correct semiclassical limit of the Master Constraint.
As a further bonus, AQG possibly can deal with topology change in the 
sense that 
it incorporates the semiclassical limits for all topologies while the 
corresponding states {\it belong to the same Hilbert space}.\\
\\
A point worthwhile noticing is that for 
convenience we used elementary variables whose classical limit coincides 
with those that are the starting point for LQG, hence our considerations 
are very much inspired by LQG. However, our purely 
combinatorial setup can be used in a much wider context, for instance it 
is conceivable that one can work with ADM variables rather than connection 
variables in the absence of fermionic matter. All one needs is to smear
the ADM variables $q_{ab},\;P^{ab}$ over regions in $\sigma$ whose 
smearing dimensions add up to three. These smearing labels are then 
promoted to elements of an abstract countable labelling set of an 
algebra whose commutation relations mimic the Poisson brackets of the 
embedded objects.\\
\\   
Much has yet to be understood about AQG. For instance, what have the exact 
solutions of the Master Constraint of AQG, when embedded, to do with the 
exact solutions of at least the spatial diffeomorphism constraint of LQG?
What we have established is that the semiclassical limit of the weighted 
square of the spatial diffeomorphism constraint agrees with the classical
generator. Hence, semiclassical states peaked on the constraint surface of 
the spatial diffeomorphism constraint are approximate solutions of the 
spatial diffeomorphism constraint\footnote{This does not contradict the 
fact that in LQG the infinitesimal generator of the diffeomorphism group 
does not exist due to the weight operator that is used in the definition 
of the square.}. But are they approximately invariant under the finite 
diffeomorphisms of $\sigma$?

Another open question is the following: Basically the master Constraint
is the weighted sum of the Hamiltonian and spatial diffeomorphism 
constraints, which when embedded look similar to the discretisations 
used in \cite{Loll}. While the Master 
constraint itself is in any case {\it non -- anomalous} we know that 
the constraints themselves do not 
close. Thus, the exact kernel of the master constraint could be empty or
may contain too few solutions because the algebra of the constraints is 
anomalous. If this is the case then, as already mentioned, one must 
modify the master Constraint. 
There are several proposals: Either one subtracts from the Master 
Constraint
the minimum of the spectrum, or one allows a whole interval of zero in the 
spectum to define solutions \cite{Klauder} or one succeeds in defining non 
-- anomalous constraints on the lattice, for instance by renormalisation 
group techniques \cite{Hasenfratz}. 

Finally, an interesting question is whether there is an algebraic 
version not only of the volume operator but also of area 
\cite{RSVol,Area}
and length operators of LQG \cite{Length}. This requires a 
diffeomorphism invariant definition of the classical version of 
these operators in terms of matter whose analytical expression uses
3d rather than 2d or 1d integrals in order that there is an embedding 
independent lift, see \cite{Complexifier} for an explanation. 
While the construction of these operators is not necessary because 
there are other functions on the classical, spatially 
diffeomorphism invariant phase space which separate the points, it 
would certainly be desirable to have those at ones disposal. We 
will leave this for future research.\\
\\ 
\\ 
\\ 
{\large Acknowledgments}\\
\\
K.G. thanks the Heinrich B\"oll Stiftung for financial support. 
This research project was supported in part by a grant from NSERC of 
Canada to the Perimeter Institute for Theoretical Physics.

\begin{appendix}

\section{Alternative Quantisation of the Spatial Diffeomorphism and 
Hamiltonian Constraints}
\label{sa}

By using the operator $Q_v$ defined in section (\ref{s2.3}) one can 
simplify the discretisations of the spatial diffeomorphism and Hamiltonian 
constraints and make the construction of the master constraint look more 
uniform. We just display the purely gravitational contributions, the 
general pattern should become clear.
\begin{itemize}
\item[C'] {\it Spatial Diffeomorphism Constraint}\\
For any $v\in V(\alpha)$ we set
\be \label{a.1}
\tilde{D}_j(v):=\frac{1}{P_v}\sum_{e_1\cap e_2=v} \frac{1}{|L(v,e_1,e_2)|} 
\sum_{\beta 
\in L(v,e_1,e_2)}
{\rm Tr}(\tau_k [A(\beta)-A(\beta)^{-1}]) E_k(e_1) E_j(e_2)
\ee 
where the sum is over unordered pairs of distinct edges adjacent to $v$ 
and where again we assumed for convenience that all edges are outgoing 
from $v$. The quantity $P_v$ is the number of contributing pairs.
\item[D'] {\it Euclidean Hamiltonian Constraint}\\
For any $v\in V(\alpha)$ we set
\be \label{a.2}
\tilde{H}_E(v):=\frac{1}{P_v}\sum_{e_1\cap e_2=v} \frac{1}{|L(v,e_1,e_2)|} 
\sum_{\beta \in L(v,e_1,e_2)}
{\rm Tr}([A(\beta)-A(\beta)^{-1}] E(e_1) E(e_2))
\ee 
where we used the same notation as above.
\item[E'] {\it Lorentzian Hamiltonian Constraint}\\
For any $v\in V(\alpha)$ we set
\be \label{a.3}
\tilde{H}(v)-\tilde{H}_E(v):=\frac{1}{P_v}\sum_{e_1\cap e_2=v}
{\rm Tr}( 
[(A(e_1) [A(e_1)^{-1},[\tilde{H}'_E,V]]),
(A(e_2) [A(e_2)^{-1},[\tilde{H}'_E,V]])]
[E(e_1), E(e_2)])
\ee
where we used the same notation as above and have set
$\tilde{H}'_E:=\sum_v [Q_v^{(1/2)}]^\dagger Q_v^{(1/2)} \tilde{H}_E(v)$.
\item[F'] {\it (Extended) Master Constraint}\\
The Extended Master Constraint is now simply given by
\ba \label{a.4}
\MC' &:=& \sum_{v\in V(\alpha)} [(Q_v^{(1/2)} G_j(v))^\dagger 
(Q_v^{(1/2)} 
G_j(v))+([Q_v^{(1/6)}]^2 \tilde{D}_j(v))^\dagger ([Q_v^{(1/6)}]^2 
\tilde{D}_j(v))
\nonumber\\
&& +
([Q_v^{(1/6)}]^2 \tilde{H}(v))^\dagger ([Q_v^{(1/6)}]^2 \tilde{H}(v))]
\ea
where appropriate coefficients are understood as in the main text in 
order to match dimensionalities and classical limit.
Not only did the constraints simplify, also all terms involved in $\MC'$
sandwich operators of the type $(Q_v^{(r)})^\dagger)^n (Q_v^{(r)})^n$. This 
is because the operators 
$D_j(v),H(v)$ transform as half -- densities when embedded, $G_j(v)$ is a 
simple density and $\tilde{D}_j(v),\tilde{H}(v)$ are double densities.  
The advantage is that the actual constraints (almost) remain polynomials
in holonomies and electric fluxes, up to appearances of the operators 
$V^(r)_v$.
\end{itemize}

\end{appendix}

\end{document}